\documentclass{ws-p8-50x6-00}

\begin{document}
%
%
\def\deg{\ifmmode ^{\rm o} \else $^{\rm o}$\fi}
\def\degC{$^{\rm o}$C}
\def\cc{\ifmmode {\rm cm}^{-3} \else cm$^{-3}$\fi}
\def\Fvar{\ifmmode F_{\rm var} \else $F_{\rm var}$\fi}
%
%
\def\arcsec{\ifmmode '' \else $''$\fi}
\def\arcmin{\ifmmode ' \else $'$\fi}
\def\arcsecpoint{\ifmmode ''\!. \else $''\!.$\fi}
\def\arcminpoint{\ifmmode '\!. \else $'\!.$\fi}
\def\farcs{\hbox{$.\!\!^{\prime\prime}$}}
%
%
\def\kms{\ifmmode {\rm km\ s}^{-1} \else km s$^{-1}$\fi}
\def\Hubble{\ifmmode {\rm km\ s}^{-1}\ {\rm Mpc}^{-1} 
	\else km s$^{-1}$ Mpc$^{-1}$\fi}
\def\ergsec{\ifmmode {\rm ergs\ s}^{-1} \else ergs s$^{-1}$\fi}
\def\ergscm2{\ifmmode {\rm ergs\ s}^{-1}\;{\rm cm}^{-2}
	  \else ergs s$^{-1}$ cm$^{-2}$\fi}
\def\efluxA{\ifmmode {\rm ergs\ s}^{-1}\;{\rm cm}^{-2}\;{\rm \AA}^{-1}
	  \else ergs s$^{-1}$ cm$^{-2}$ \AA$^{-1}$\fi}
\def\efluxHz{\ifmmode {\rm ergs\ s}^{-1}\;{\rm cm}^{-2}\;{\rm Hz}^{-1}
	  \else ergs s$^{-1}$ cm$^{-2}$ Hz$^{-1}$\fi}
%
%
\def\Msun{\ifmmode {\rm M}_{\odot} \else ${\rm M}_{\odot}$\fi}
\def\Lsun{\ifmmode {\rm L}_{\odot} \else ${\rm L}_{\odot}$\fi}
%
%
\def\qo{\ifmmode q_{0} \else $q_{0}$\fi}
\def\Ho{\ifmmode H_{0} \else $H_{0}$\fi}
\def\ho{\ifmmode h_{0} \else $h_{0}$\fi}
\def\qo{\ifmmode q_{0} \else $q_{0}$\fi}
\def\ao{\ifmmode a_{0} \else $a_{0}$\fi}
\def\to{\ifmmode t_{0} \else $t_{0}$\fi}
%
%
\def\ltsim{\raisebox{-.5ex}{$\;\stackrel{<}{\sim}\;$}}
\def\gtsim{\raisebox{-.5ex}{$\;\stackrel{>}{\sim}\;$}}
\def\vFWHM{\ifmmode V_{\mbox{\tiny FWHM}} \else
            $V_{\mbox{\tiny FWHM}}$\fi}
%
%
\def\Halpha{\ifmmode {\rm H}\alpha \else H$\alpha$\fi}
\def\Hbeta{\ifmmode {\rm H}\beta \else H$\beta$\fi}
\def\Hgamma{\ifmmode {\rm H}\gamma \else H$\gamma$\fi}
\def\Hdelta{\ifmmode {\rm H}\delta \else H$\delta$\fi}
\def\Lya{\ifmmode {\rm Ly}\alpha \else Ly$\alpha$\fi}
\def\Lyb{\ifmmode {\rm Ly}\beta \else Ly$\beta$\fi}
\def\hi{H\,{\sc i}}
\def\hii{H\,{\sc ii}}
\def\4686{He\,{\sc ii}\,$\lambda4686$}
\def\hei{He\,{\sc i}}
\def\heii{He\,{\sc ii}}
\def\ci{C\,{\sc i}}
\def\cii{C\,{\sc ii}}
\def\ciii{\ifmmode {\rm C}\,{\sc iii} \else C\,{\sc iii}\fi}
\def\civ{\ifmmode {\rm C}\,{\sc iv} \else C\,{\sc iv}\fi}
\def\ni{N\,{\sc i}}
\def\nii{N\,{\sc ii}}
\def\niii{N\,{\sc iii}}
\def\niv{N\,{\sc iv}}
\def\nv{N\,{\sc v}}
\def\oi{O\,{\sc i}}
\def\oii{O\,{\sc ii}}
\def\oiii{O\,{\sc iii}}
\def\oiv{O\,{\sc iv}}
\def\ov{O\,{\sc v}}
\def\ovi{O\,{\sc vi}}
\def\nev{Ne\,{\sc v}}
\def\mgi{Mg\,{\sc i}}
\def\mgii{Mg\,{\sc ii}}
\def\siIV{Si\,{\sc iv}}
\def\si{S\,{\sc i}}
\def\sii{S\,{\sc ii}}
\def\siii{S\,{\sc iii}}
\def\caii{Ca\,{\sc ii}}
\def\feii{Fe\,{\sc ii}}
\def\aliii{Al\,{\sc iii}}
\def\o5007{[O\,{\sc iii}]\,$\lambda5007$}

\title{Variability of Active Galactic Nuclei}

\author{Bradley M. Peterson}

\address{Department of Astronomy, The Ohio State University,
140 West 18th Avenue, Columbus, OH 43210, 
USA\\E-mail: peterson@astronomy.ohio-state.edu}


\maketitle

\abstracts{
Continuum and emission-line variability of active galactic
nuclei provides a powerful probe of  
microarcsecond scale structures in the central regions of
these sources. In this contribution, we review
basic concepts and methodologies used in analyzing
AGN variability. We develop from first principles
the basics of reverberation mapping, and pay special
attention to emission-line transfer functions. We
discuss application of cross-correlation analysis
to AGN light curves. Finally, we provide a short
review of recent important results in the field.}

\section{Introduction}
\label{sect:intro}
The study of
multiwavelength variability of active galactic nuclei (AGNs) is
now a major branch of the field, enabled largely by
the availability of suitable facilities for
long-term studies of faint sources at many wavelengths;
some of the scientific arguments for development of
these facilities have been based on exactly these
programs, which are now leading to improved understanding
of the AGN phenomenon. It is only within the last
few years that the evidence for supermassive black holes
in both active and non-active galaxies has gone from
circumstantial to compelling, and the potentially
most powerful technique for measuring black-hole
masses in AGNs is through study of broad emission-line
variability. The existence of accretion disks in
AGNs is still far from proven, but the evidence for
them is improving, again as a result of variability
studies. Multiwavelength monitoring observations are
beginning to show the relationships between
variability in different bands, and the hope is that
once the phenomenology is better known,
our understanding of the physics will follow.

In this contribution, we will cover the basic characteristics
of AGN variability 
and provide what we hope is some relevant historical background.
Because of the current
importance of emission-line variability studies, we
will develop the theory of reverberation mapping from
first principles. One of the most powerful and widely
used tools in the analysis of emission-line and continuum
variability data is the technique of cross-correlation, and we
will therefore describe in some detail the application of this method
to AGN data. Throughout this contribution, we will
concentrate almost exclusively on non-blazar AGNs,
those for which we believe most of the observed UV/optical
emission originates in an accretion disk rather than
in a relativistic jet: many of the techniques described
here are also applicable to blazars, however.
For a fairly recent discussion of blazar variability
results, we refer the reader to the fine review
by Ulrich et al.\cite{UMU97}.
We emphasize that we intend for this contribution
to be primarily instructional; this should not be
misconstrued as a comprehensive review of the state of the field.
Our intent is to provide both students and researchers
who already have some familiarity with AGNs with 
enough background to read critically the current literature
on AGN variability and understand the strengths and
weaknesses of the method.

\section{Background and Basic Phenomenology}
\label{sect:back}

AGNs show flux variations over
the entire electromagnetic spectrum. Indeed, variability
was one of the first recognized properties of quasars\cite{MaSa63}$^{,}$\cite{SmHo63}.
Early investigations established that significant 
variations ($\gtsim0.1$\,mag)
in the optical brightness of quasars
could occur on time scales as short as days. 

Detection of rapid variability in quasars was
a remarkable discovery at the time because it 
implies that the size of the continuum-emitting region must be
of 
order light days ($1\ \mbox{\rm lt-day} = 2.6 \times 10^{15}$\,cm),
based on source coherence arguments: for a source 
to vary coherently, the entire emitting region must be causally 
connected, which implies a maximum size for the source
based on light-travel time. Suppose, for example, 
that the brightness of the source doubles in a week;
we can immediately conclude that the emitting region
must be no larger than one light week in radius
on the basis of causality.
We might suppose that in fact there are multiple
emitting regions varying at random, but the number of
such regions must be limited or the stochastic variations
would be averaged out. Individually, then, the
purported independent regions face a similar size
limit from causality, and the conclusion that
the emitting regions are very small cannot be
avoided. Historically, this is the quasar problem:
how can so much energy, the equivalent of as much as trillions 
of stars, be produced in a region that is about
the size of the Solar System?

The detection of quasar variability was a critical part
of the argument that AGNs are powered by supermassive 
black holes. The original arguments for supermassive
black holes in AGNs were based on mass constraints
from the Eddington limit and size constraints from
variability\cite{Pe97}. 
The Eddington limit is the requirement
that gravitational forces on an ionized gas exceed
outward radiation pressure, which translates to 
a requirement that 
\begin{equation}
M \ge M_{\rm Edd} \equiv 8 \times 10^5 
\left(\frac{L}{10^{44}\ {\rm \ergsec}} \right)\ \Msun \mbox{\ \ .}
\end{equation}
Rapid variations, in some cases on time scales as
short as a day, require an emitting region less than
a light day in radius, which for a $10^{44}$\,\ergsec\
AGN corresponds to $\sim10^4 R_{\rm grav}$, where
$R_{\rm grav} = GM/c^2$ is the gravitational radius.
Later arguments for supermassive black holes in AGNs
focussed on how to power AGNs by gravitational
accretion\cite{Sh78}. The deep gravitational potential leads to 
an accretion disk that radiates most strongly across the
UV/optical spectrum,
and for AGN masses above the Eddington limit,
the thermal emission should peak in the near UV.
Indeed, then, the broad UV/optical feature known as the
``big blue bump'' can plausibly be identified with accretion-disk emission. 
Furthermore, the intense magnetic fields expected in disks
could provide a mechanism for jet collimation.

\subsection{Basic Characteristics of Variability}

AGNs have been found to be variable at all wavelengths
at which they have been observed. The variations appear to be
aperiodic and have variable amplitude. While variability in
high-luminosity AGNs (quasars) 
was reported soon after their discovery, variability in 
lower-luminosity AGNs
(Seyfert galaxies) was not reported\cite{FPW67} until 1967,
and was less dramatic. 
The reason for this is probably quite simple:
most of the quasars that were monitored are now known to 
be the jet-dominated sources known as ``blazars''. i.e., BL Lac 
objects and optically violent variables (OVVs). The optical
identifications of quasars were based on coincidence with
radio-source positions, which naturally led to biases towards
radio-loud quasars, blazars in particular.
Indeed, the original arguments about size and
variability time scales in retrospect apply to the jets,
not necessarily what we now identify as thermally emitting
accretion disks.  Nevertheless, 
the original conclusions about AGN sizes 
proved to be generally correct for both blazars and non-blazars.

\begin{figure}
\begin{center}
    \leavevmode
  \centerline{\epsfig{file=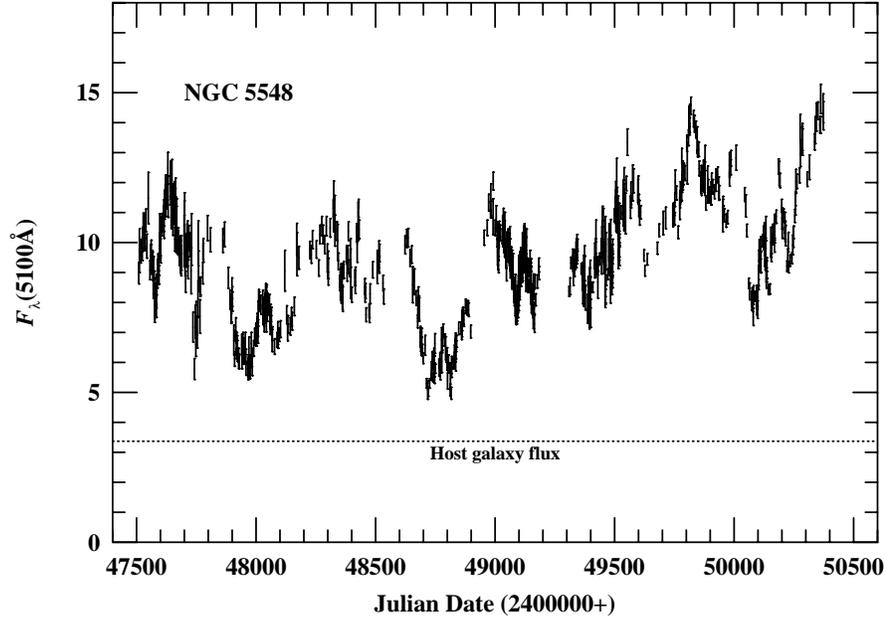,width=11.8cm,angle=0}}
  \end{center}
\caption{Optical (5100\,\AA) light curve of NGC 5548 from
late 1988 to late 1996. The horizontal line indicates
the constant contribution from starlight through the
standard aperture used, which projects to $5''\times
7''\!.5$. From Peterson et al.$^{71}$ \copyright 1999 AAS.
\label{fig:ngc5548lc}}
\end{figure}

\bigskip
\noindent
{\em UV/Optical Variability.}
Fig.~\ref{fig:ngc5548lc}  shows a light curve for 
a typical Seyfert 1 galaxy,
NGC 5548, which will serve as a continuing example through
this chapter as it is one of the best-studied objects of this class.
While no periodic behavior has been identified,
there are some basic parameterizations
that allow us to characterize the variability.
An example is shown in Fig.~\ref{fig:deltaF}. We compare
each flux measurement at an arbitrary time $t_i$ 
with the flux at every later time $t_j$. The quantity
we show is the flux ratio as a function of
the time interval between observations 
$\Delta t = t_j - t_i$, i.e.,
$\Delta \log F = \log F(t_j)/F(t_i)$, where
in each case the contaminating flux due to starlight
in the host galaxy (as shown in Fig.~\ref{fig:ngc5548lc}) 
has been first 
removed. Fig.~\ref{fig:deltaF} shows that NGC 5548 shows little
variability on time scales shorter than a few days,
but on time scales of several weeks or months, very
large variations can be observed. We note in passing
that the quantity shown here is closely related to 
the ``structure function''.
The structure function is simply the mean absolute
value of $2.5 \Delta \log F(\Delta t)$, i.e., the
mean difference in magnitudes between observations
separated by times $\Delta t$.

\begin{figure}
\begin{center}
    \leavevmode
\vspace{4cm}
  \end{center}
\caption{Difference in optical flux as a function of 
interval between observations $\Delta t$ for the light
curve shown in Fig.~\ref{fig:ngc5548lc}. The starlight contribution
as shown in Fig.~\ref{fig:ngc5548lc} has been subtracted off.
\label{fig:deltaF}}
\end{figure}

\begin{figure}
\begin{center}
    \leavevmode
  \centerline{\epsfig{file=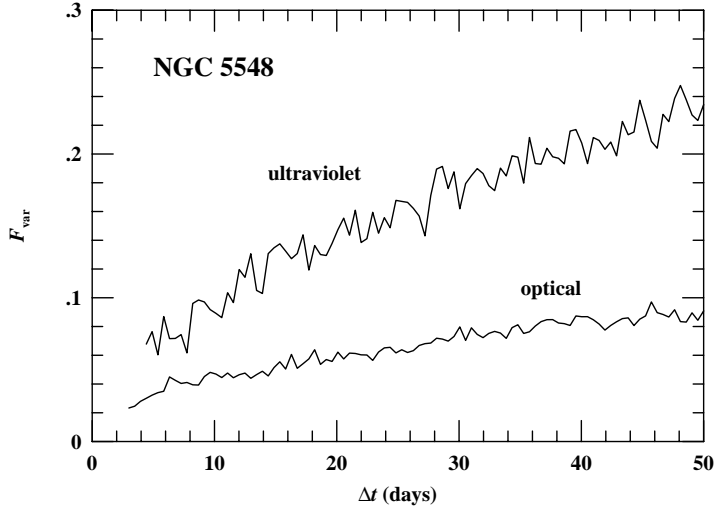,width=9.5cm,angle=0}}
  \end{center}
\caption{Variability parameter \Fvar\ from ultraviolet
and optical continuum measurements of NGC 5548, as a
function of interval between observations;
the optical data used are those shown in 
Fig.~\ref{fig:ngc5548lc}. Courtesy of S.\ Collier.
\label{fig:Fvar}}
\end{figure}

A common parameter to characterize variability is
the mean fractional variation,
\begin{equation}
\label{eq:Fvar}
\Fvar = \frac{ \sqrt{\sigma^2 - \delta^2}}
{\left< f \right> } \mbox{\ \ ,}
\end{equation}
where the quantities are (a) the mean flux
for all $N$ observations,
\begin{equation}
\left< f \right> = \sum_{i=1}^{N} f_i \mbox{\ \ ,}
\end{equation}
(b) the variance of the flux (as observed),
\begin{equation}
\sigma^2 = \frac{1}{N}
\sum_{i=1}^{N} \left( f_i - \left< f \right> \right)^2 \mbox{\ \ ,}
\end{equation}
and (c) the mean square uncertainty of the fluxes
\begin{equation}
\delta^2 = \frac{1}{N}\sum_{i=1}^{N} \delta_i^2 \mbox{\ \ .}
\end{equation}
Even if the continuum is constant, there will still be
apparent flux variations simply due to measurement uncertainties
(noise); the virtue of \Fvar\ is that it adjusts the
fractional variation downwards to account for the
effect of random errors. The parameter \Fvar\ is thus
sometimes referred to as the ``excess variance.''
In Fig.~\ref{fig:Fvar}, we show \Fvar\ as a function of time
interval UV and optical variations in 
NGC 5548. 
This shows that UV continuum variations are typically
around 10--20\% on time scales of about a month.
The fractional variations in the optical are less pronounced,
at least in part because of contamination of the fluxes
by a large constant contribution from starlight in the
host galaxy.

\bigskip
\noindent
{\em X-Ray Variability.}
Rapid X-ray variability is a hallmark of AGNs
(for a review, see Mushotzky et al.\cite{MDP93}).
It is of historical importance since it
effectively eliminates alternatives to supermassive black 
holes, e.g., massive stars or starbursts, as competing
explanations for the high luminosities of AGNs.
X-rays are expected to arise near the event horizon
of the black hole, so the shortest time-scale X-ray variability 
is expected on a few times the crossing time
\begin{equation}
\label{eq:crossing}
t_{\rm crossing} = \frac{R_{\rm grav}}{c} =
10 \left( \frac{M}{10^6\ \Msun} \right)\ {\rm s.}
\end{equation}

Prior to the 1990s and the advent of RXTE, the
best  X-ray monitoring data was from EXOSAT,
which was in a  high-Earth orbit that allowed up to 80 hours of 
uninterrupted observations. The
EXOSAT ``long-looks'' at variable AGNs established that variability is most 
rapid in low-luminosity systems.
 
A useful way to characterize variability is in terms of 
the ``power density spectrum'' (PDS), which is the product of 
the Fourier transform of the light curve and its complex conjugate.
The PDS for AGNs is often parameterized as a power law,
\begin{equation} 
\label{eq:PDS}
P(f) = f^{-\alpha} \mbox{\ \ .} 
\end{equation}
The EXOSAT data\cite{Mc88} showed that
AGN X-ray variations can be characterized by PDS indices in
the range $1 \ltsim \alpha \ltsim 2$ over time scales of
hours to months. The total power in the 
variations is given by integrating the PDS  over all frequencies.
Thus, the  PDS must turn over at low frequencies 
(i.e., $\alpha$ as defined above must become less than unity)
to prevent divergence in the total power. Such breaks in the
PDS are observed  in stellar-mass X-ray sources, and the
turnover frequency correlates inversely with mass, though
the fundamental reason for this is not understood.
The basic idea is that the mass
of the black hole can be inferred, since 
$M_{\rm BH} \propto R_{\rm grav} \propto t_{\rm crossing}$.
If we scale AGNs relative to stellar-mass systems 
(which have turnover frequencies $f \approx 0.1$\,Hz), 
we expect that the turnover frequencies for AGNs will occur at 
about $10^{-8}$\,Hz. 
In only one case, NGC~3516, has there been a plausible
detection of the turnover frequency. Edelson \& Nandra\cite{EdNa99}
find that for this Seyfert 1 galaxy the turnover frequency
is $f \approx 4 \times 10^{-7}$\,Hz, which corresponds to 
a time scale of about one month. The mass inferred, again
scaling relative to stellar-mass systems, is in the range
$10^6$--$10^7\,\Msun$.

Periodicities in X-ray light curves, which might reflect
orbital or precession periods, have been searched for, but never found.
As a historical footnote, however, it is worth mentioning that 
one such detection was claimed\cite{MiBr89},
namely a 12,000~s period in NGC~6814. 
However, ROSAT
observations revealed that the variable source is in fact 
a foreground Galactic binary in the 
same field as the AGN\cite{Ma_ea93}.

\subsection{Origin of the Variations}

At a fundamental level, the physical origin of variations
is not known, although accretion-disk instabilities are probably
involved. For example, Kawaguchi et al.\cite{Ka_ea00} show
that continuum variations with a PDS of the form of Eq.~(\ref{eq:PDS})
can be explained by magnetohydrodynamic instabilities,
specifically disconnection events, within the disk.
Variable accretion rates have also been considered.
In some specific atypical cases, variations have been attributed
to variable obscuration of the nuclear source and to  microlensing
due to stars in the host galaxy.

There are a number of important physical time scales that
might be associated with variability. We include them
here in the convenient form given by Edelson \& Nandra\cite{EdNa99}.
The first of these
is the crossing time already mentioned in 
Eq.~(\ref{eq:crossing}), which we rewrite here as
\begin{equation}
\label{eq:scalecross}
t_{\rm crossing} = 0.011 M_7 \left( \frac{r}{10 R_{\rm grav}} 
\right)\ {\rm days},
\end{equation}
which is the time it takes a radiative signal to cross the
X-ray emitting region (assumed to be at $r \ltsim 10 R_{\rm grav}$).
Here $M_7$ is the black-hole mass in units of $10^7$\,\Msun.
Variations might also be expected on the time scale of the
orbital period,
\begin{equation}
\label{eq:orbital}
t_{\rm orbital} = 0.33 M_7 \left( \frac{r}{10 R_{\rm grav}} 
\right)^{3/2}\ {\rm days}.
\end{equation}
Thermal instabilities might also cause variations
on the time scale for their development,
\begin{equation}
\label{eq:thermal}
t_{\rm thermal} = 5.3 \left( \frac{\alpha}{0.01} \right)^{-1}
M_7 \left( \frac{r}{10 R_{\rm grav}} 
\right)^{3/2}\ {\rm days},
\end{equation}
where $\alpha$ is the viscosity parameter.
Mechanical instabilities may propagate as acoustic waves,
which will travel at the sound speed and thus cross the disk on
a time scale
\begin{equation}
\label{eq:sound}
t_{\rm sound} = 33 \left( \frac{r}{100H} \right)
M_7 \left( \frac{r}{10 R_{\rm grav}} 
\right)^{3/2}\ {\rm days},
\end{equation}
where $H$ is the disk thickness. And finally,
the time scale over which the effects of variations in
the accretion rate will propagate through the disk is
given by the drift speed,
\begin{equation}
\label{eq:drift}
t_{\rm drift} = 53000 \left( \frac{r}{100H} \right)^2
\left( \frac{\alpha}{0.01} \right)^{-1}
M_7 \left( \frac{r}{10 R_{\rm grav}} 
\right)^{3/2}\ {\rm days}.
\end{equation}

\subsection{Blazar Variability}
Variability properties of blazars are distinct from those 
of other non-beamed AGNs and should be mentioned separately.
Blazars are characterized by extreme variability
at all wavelengths. Unlike radio-quiet quasars or
Seyfert galaxies, significant infrared and radio continuum variability
is observed in blazars. Furthermore, the polarization
of the continuum radiation is also significant
(i.e., greater than a few percent), and the 
degree of polarization and amplitude of variability 
correlate with luminosity, which is the opposite case for non-blazars.
Blazars are also the only sources detected at TeV energies,
and the TeV fluxes can vary by as much as a factor of 10 in one day.
All of these properties indicate that the continuum is 
dominated by emission from relativistic jets, as these
characteristics suggest a non-thermal (synchrotron or
inverse Compton) origin.
 
\subsection{Emission-Line Variability}
The broad emission lines in AGN spectra can vary 
both in flux and in profile. Over time scales of months
and years, the changes can be very dramatic, but on
shorter time scales they are more subtle. The first detection
of emission-line variations was by Andrillat \& Souffrin\cite{AnSo68},
based on photographic spectra of the Seyfert 1
galaxy NGC 3516. There were a few subsequent 
reports\cite{ToOs76}$^,$\cite{Ph78}, but these
cases seemed to be widely regarded as ``curiosities'' that
did not generate much follow-up work. The basic problem was
that only very large changes could be detected photographically 
or with the intensified television-type scanners 
that were commonly used in AGN spectroscopy from the
mid-1970s to mid-1980s. In the few cases where
clear variations were detected, the changes
were often dramatic; there were sometimes claims of Seyferts changing ``type'' 
as broad components of emission lines appeared or disappeared.

Pronounced variability of broad emission-line profiles was detected 
in the early 1980s by a number of investigators.
Profile variations were originally thought to be due to excitation
inhomogeneities; excitation pulses propagating through a 
broad-line region (BLR)
with an ordered velocity field would produce 
features
that could propagate across the profile with time.
This concept led to the development of reverberation mapping,
which is described in detail in Sec.~{\ref{sect:rmtheory}}.
Peterson\cite{Pe88} 
reviews early work on emission-line variability.

\begin{figure}
\begin{center}
    \leavevmode
  \centerline{\epsfig{file=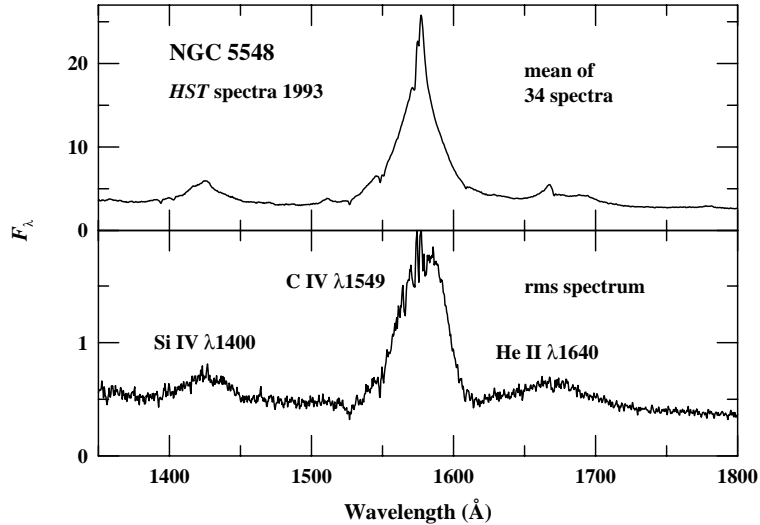,width=10cm,angle=0}}
  \end{center}
\caption{The top panel shows the mean spectrum computed
from 34 HST spectra of the variable Seyfert 1 galaxy NGC 5548$^{40}$.
The lower panel shows the rms
spectrum based on variations around this mean. The rms
spectrum thus isolates the variable components of the
spectrum. Fluxes are in units of 
$10^{-15}$\,\efluxA.
\label{fig:ngc5548rms}}
\end{figure}

A useful way to isolate the variable part of an emission
line is shown in Fig.~\ref{fig:ngc5548rms}.
The upper panel shows a mean spectrum formed from 
34 individual HST spectra of NGC 5548. 
The lower panel shows the ``root-mean square'' (rms) spectrum 
which is formed from the same data simply by computing
the rms flux at each wavelength. Constant
features, such as narrow emission lines and
host-galaxy flux, do not appear in the rms spectrum.

\subsection{The First Monitoring Programs}
The early 1980s saw the first attempts to monitor the
UV/optical continuum and emission-line variations in
Seyfert 1 galaxies. There were two reasons this happened
when it did. First, there was a realization that
variability afforded a powerful 
probe of the structure and kinematics of AGNs on projected scales
of microarcseconds. Variability was recognized as
an important new tool with which to
study enigmatic quasars. Second, the right technology
for such investigations became widely available:
it was possible to attempt such  programs on account of 
(a) IUE,
which for the first time allowed precision UV spectroscopy
of low-redshift extragalactic objects, and (b) the
proliferation of linear electronic detectors 
(first Reticons and Image Dissector Scanners, and later
CCDs) on moderate-size (1--2m) ground-based telescopes.

One of the first significant monitoring programs was
a multiple-year  IUE-based program on  NGC 4151, which
was carried out by a European consortium led by
M.V.\ Penston and M.-H.\ Ulrich\cite{Ul_ea91}. Ultraviolet spectra
were obtained with a typical sampling interval of 2--3 months. 
The program showed that variations in the UV and optical
continua were closely coupled. It also revealed  that the
emission-line flux variations are correlated with continuum
variations, but that different lines respond in different ways,
both in amplitude and in time scale. These data also showed
a complicated relationship between UV and X-ray variations
and led to the discovery of variable absorption lines
in the ultraviolet.

The galaxy  NGC 4151 was also monitored spectroscopically
in the optical at Lick Observatory by Antonucci \& Cohen\cite{AnCo83}.
They found that the Balmer lines seemed to respond to continuum
variations on a time scale less than around one month
(their typical sampling interval). They also found that
relative to \Hbeta\ and \Halpha, the higher-order Balmer 
lines and \4686\ varied with higher amplitudes.

Arakelian (Akn) 120 was the first higher-luminosity Seyfert
that was monitored fairly extensively in the 
optical\cite{Pe_ea83}$^,$\cite{Pe_ea85}
as a result of dramatic
Balmer-line profile changes that had been detected 
earlier\cite{Fo_ea81}$^,$\cite{Ko_ea81}$^,$\cite{ScRa81}.
Peterson et al.\cite{Pe_ea85}
found that the time scale for the response of 
\Hbeta\ to continuum variations suggested a BLR size of less than 1 light 
month across. This was a surprising result as it 
suggested that there was a serious problem with existing estimates of 
sizes of the BLR that were based on photoionization equilibrium
modeling, as these indicated the BLR should be about an order-of-magnitude
larger than this. The upper limit on the BLR size was similar to
that obtained by Antonucci \& Cohen\cite{AnCo83} for NGC 4151, but
because Akn 120 is a higher-luminosity source, 
the monthly sampled data provided a more critical challenge to BLR models.

Not surprisingly, the results from these earlier monitoring
programs were controversial. 
Several observational problems could be identified:
\begin{enumerate}
\item {\em Undersampling of the variations.}
The variations tended to be undersampled because 
the original programs for monitoring 
Seyfert galaxies were designed for BLRs that 
were thought to be many light months in size.
For example, in the case of Akn 120, Peterson et al.\cite{Pe_ea85}
were looking for profile structures that were expected
to cross the line profile on a time scale of a year or
so, and monthly observations should have been sufficient to carry out
this program. There was certainly at the time no reason
to believe that higher sampling rates should be required;
indeed, proposals to observe AGNs as often as once per month 
were sometimes deemed to be ``oversampled'' by 
telescope allocation committees! There is no
obvious algorithm to determine whether or not the variations 
that have been observed are undersampled, but it is
quite obvious that if the results depend on individual data points, 
the light curve is almost certainly undersampled and
any conclusions drawn must be eyed with suspicion.
A very simple operational criterion for adequate sampling 
is that if the results do not change much when individual points are 
removed from the light curve, the light curve is
probably not seriously undersampled. A nice simple test is to
divide a light curve into two parts, one comprised of the even-numbered
points (i.e., second, fourth, 
etc., in the time-ordered
series) and one comprised of the odd-numbered points. If
the two light curves are still very similar, i.e., the
important features appear in both light curves, then the original
light curve is probably adequately sampled.
\item {\em Low S/N of the light curves.}
If the detected variations are not large compared to the 
signal-to-noise ratio (S/N) of typical data, then
spurious results can be obtained. Stated another way,
$\Fvar$ (Eq.~(\ref{eq:Fvar})) must be significantly greater than zero.
This was a serious 
issue in the case of some of the earlier data obtained with image dissector 
scanners and Reticon arrays, for which 
uncertainties in AGN line and continuum fluxes
were typically around 8--10\%. To a large degree, this
problem has been obviated by use of CCDs, for which
typical errors in the 1--3\% range are routinely achieved.
\item {\em Systematic errors.}
There are two sometimes-related types of insidious errors that can adversely
affect
time series analysis: (a) correlated continuum/line errors, and
(b) aperture effects. 

Correlated errors are due to systematic
flux-calibration errors. Basically, if the flux calibration of
a spectrum is incorrect, both the continuum and emission-line
fluxes measured from it will be in error in the same sense;
if the calibration is too high, both the emission-line and
continuum fluxes will be too high. This introduces an artificial
correlation between the line and continuum at zero lag,
and can thus bias the measurement of the true lag between them
to artificially small values. 

Aperture effects occur when the amount of flux entering
a spectrograph is not fixed,  on account of pointing
or guiding errors, or variations in seeing in the case
of ground-based observations. 
In point-like sources like stars, this
affects only the overall photometric accuracy. In nearby
AGNs, however, both the narrow-line region (NLR) and host-galaxy are
spatially resolved, and the aperture geometry,
centering and guiding, and seeing variations can lead
to apparent spectral variations. Depending on their
nature, aperture effects can cause either correlated, 
uncorrelated, or even anticorrelated errors in the line
and continuum fluxes. 
\end{enumerate}

\subsection{Spectrophotometric Flux Calibration}
In this section, we will outline some of the important
considerations for flux calibration of spectroscopic
monitoring data, directed primarily towards ground-based
optical observers.

In AGNs, aperture effects arise because the source
is comprised of multiple components, some of which have
angular structure on scales similar to the width
of the point-spread function (PSF). 
The basic requirement for accurately flux-calibrating
AGN spectra is that stable fractions of light from the AGN
continuum source and BLR (both point-like at even
the 0\farcs01 level) and the NLR and
the host galaxy (both extended even at arcsecond levels) must 
enter the aperture. We note that even in the case of space-based
observations, there is a trade-off between aperture size and 
pointing uncertainty: the goal is to minimize the amount of host-galaxy
starlight entering the aperture (arguing for a smaller aperture),
while ensuring that the amount of admitted starlight is constant
(arguing for an aperture large relative to the pointing accuracy
of the telescope). In the
UV, however, this is a less-significant problem because the host 
galaxy is so much fainter than the AGN itself.

Mitigation of aperture effects is one of the most
important considerations for a ground-based monitoring program.
This is important to keep in mind, as most
AGN observers are used to background-limited 
observations of point sources, and this calls for adjusting 
the slit width for variations in seeing in order 
to optimize the S/N of the data. However,
{\em this is exactly the wrong thing to do if you are 
monitoring Seyfert galaxy variations}, since calibration 
accuracy is almost always determined by 
systematics rather than photon statistics.
Most observers {\em do} know that you need to open up the 
aperture for absolute spectrophotometry, however, and
this is precisely what needs to be done in monitoring
programs, as we will explain below.

The standard method of absolute flux calibration is to
determine from observations of spectrophotometric standard
stars how counts per second per pixel on the detector
translates to flux per unit wavelength. However, 
standard absolute spectrophotometry is far too 
inaccurate for ground-based AGN monitoring; the typical accuracy that can be
achieved on photometric nights is about 5\%. Furthermore, at most
observing sites, only a relatively small fraction of nights
are of sufficient quality and stability for absolute
spectrophotometry to be useful.
Even at a good site like Kitt Peak, fewer than 1/3 of 
the nights are photometric.  However, more 
than 2/3 of all nights (including bright time) are 
good enough to get quality spectra of bright AGNs 
with a 1--2-m telescope with a typical photometric accuracy 
about 15\%. Since we need a high rate of sampling and
do not wish to discard more than half of the potentially
usable nights, secondary higher-accuracy calibration methods must 
be used. In ground-based AGN monitoring, there are two commonly
used high-accuracy flux-calibration methods, which we outline
below:
\begin{enumerate}
\item {\em Relative spectrophotometry.} This method involves
simultaneous observation of a nearby field star
(that must be shown to be non-variable!) in long-slit mode.
Absolute calibration can be achieved by calibrating 
spectra of the comparison star in the usual fashion
described above on nights of photometric quality.
This method has been used to attain internal (relative)
accuracies of $\sim1$--2\%\cite{Co_ea98}$^,$\cite{Kas_ea00}. 
The limitations of this method
are (a) an extremely accurate slit response function
(i.e., relative sensitivity as a function of position
in the slit) is required, (b) a large slit and high pointing
and guiding accuracy is required to avoid aperture/seeing effects,
and (c) this method is difficult to use shortward of 
$\sim4000$\,\AA\ because the requirement of fixed position angle 
(to observe the AGN and comparison star simultaneously)
leads to atmospheric dispersion problems.
\item {\em Internal calibration.} This is based on using
a constant-flux component in the spectrum (e.g., narrow emission lines)
to refine  absolute calibration. In this way, internal
accuracies of $\sim2$\% have been routinely achieved\cite{Pe_ea98a}$^,$
\cite{Pe_ea99}.
The [\oiii]\,$\lambda\lambda4959$, 5007 doublet is excellent for
this method because these lines are strong, relatively unblended, and
close in wavelength to \Hbeta\ and \4686. Unlike the broad
lines, the narrow lines are almost always expected to be constant
in flux. The light-crossing time for the NLR is
large (typically 100--1000 years) as is the
recombination time (about 100 years; see Eq.~(\ref{eq:rectime})
in Sec.~{\ref{sect:rmtheory}}),
so any short-term variability is smeared out. 
This method also has limitations:
(a) most narrow lines are blended and/or weak,
(b) the method works best at reasonably high spectral resolution,
when the narrow-lines are at least partially resolved,
and (c) in nearby AGNs, the NLR is often extended on scales of
arcseconds, which introduces the possibility of aperture effects.
\end{enumerate}

The last of these problems deserves serious consideration.
If the NLR is extended, measuring a consistent flux is 
difficult due to aperture effects, as described
above; either centering and guiding errors or differences
in the seeing between observations can alter the
calibration of the spectrum. However, it is possible
to understand the nature of these effects so that
in principle one can correct for them, or at minimum
one can pursue mitigation strategies to reduce their influence.
To do this, one must know (a) the surface brightness
distributions of the extended components (the host
galaxy and the NLR) and (b) the PSF, including the
seeing component, for the instrument at the time of
the observations. 

\begin{figure}
\begin{center}
    \leavevmode
  \centerline{\epsfig{file=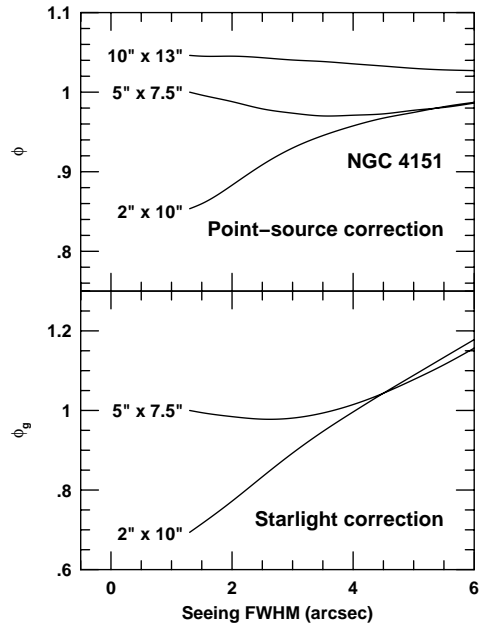,width=6.3cm,angle=0}}
  \end{center}
\caption{Aperture corrections (as defined by Peterson et al.$^{68}$) 
for NGC 4151. These are multiplicative adjustments
that must be made to (a) the broad emission-line and
AGN continuum fluxes (top panel) and (b) the host-galaxy
starlight contribution (bottom panel) as a function of
seeing when the spectra are calibrated by assuming a
constant flux in [\oiii]\,$\lambda\lambda 4959$, 5007.
The reference point is a $5''\times7\farcs5$ aperture
and seeing ${\rm FWHM} = 1\farcs3$. 
For sufficiently large apertures, the correction is negligible
under even quite poor observing conditions; for small apertures,
seeing effects can produce apparent variability.
The starlight correction factor for the $10''\times13''$ 
aperture has a value of $\sim1.6$ and varies only weakly
with seeing.
\label{fig:ngc4151cor}}
\end{figure}

With a good model for the surface brightness distributions,
we can still use the narrow-lines for flux calibration,
and then apply a seeing-dependent flux 
correction\cite{Wa_ea92}$^,$\cite{Pe_ea95}.
Even
if the seeing is not measured, such models can be
used to give us an idea of how badly our observations
might be adversely affected by seeing variations.
As an example, the corrections to the point-source fluxes
(broad lines and AGN continuum) and host galaxy
flux that need to be applied in the
case of NGC 4151 are shown in Fig.~\ref{fig:ngc4151cor} for different
aperture geometries. We can readily conclude from these
diagrams that seeing-dependence of the correction
factors is smaller for larger apertures, as can be
expected intuitively.
A useful generalization is that the corrections 
for seeing effects are generally not necessary 
if the aperture in both dimensions is larger than the 
worst seeing, as measured by the full-width at half-maximum
(FWHM) of a stellar image.

\subsection{Requirements for a Successful Monitoring Program}
In order to avoid misleading systematic errors of the 
sort discussed thus far, care must be taken in designing
an observational monitoring program. The fairly obvious basic requirements
that we should keep in mind are:
\begin{enumerate}
\item {\em The monitoring program must have sufficient sampling.}
Specifically, the observations should be closely spaced  
in time relative to physical time scales of interest. In terms of
total duration, the program should be at least three times as long as 
the longest time scale to be investigated.
\item {\em The S/N of the light curve must be high enough 
to detect variations on the shortest interesting physical 
time scale.} Inspection of Fig.~\ref{fig:deltaF}, for example, demonstrates
that if you can measure continuum fluxes to only 10\% accuracy,
3$\sigma$ detections of continuum variability in Seyfert 1
galaxies will be extremely rare on time scales as long as months.
\item {\em Systematic effects must be obviated.} The data must
be as homogeneous as possible. The easiest way
to do this is to acquire the data with a single instrument in
a stable configuration that is suitable for all 
the observing conditions expected over the duration of the program.
\end{enumerate}

\section{Theory of Reverberation Mapping}
\label{sect:rmtheory}

After emission-line variability was detected, it became
clear to a number of 
investigators\cite{BKS72}$^,$\cite{Fa80}$^,$\cite{BlMc82}$^,$\cite{CFP82}$^,$\cite{AnBo83}
that the
kinematics and the geometry of the BLR can 
be tightly constrained by characterizing the emission-line response to 
continuum variations. The time delay between continuum and
emission-line variations are ascribed to 
light travel-time effects within the BLR; the
emission lines ``echo'' or ``reverberate'' to the continuum changes.
The Blandford \& McKee paper\cite{BlMc82},  regarded as the seminal
paper in the field, first introduced the term
``reverberation mapping'' to describe this process.
Reviews of progress in reverberation mapping are provided
by Peterson\cite{Pe93} and Netzer \& Peterson\cite{NePe97}.

Because the emission lines vary with the UV/optical continuum 
(with some small time delay), we can immediately draw several
important conclusions:
\begin{enumerate}
\item The line-emitting clouds are close to the continuum source,
i.e., the light-travel time across the BLR is small.
\item The line-emitting clouds are optically thick. If they
were optically thin, the line emission from them would change
little as the continuum varied.
\item The observable UV/optical continuum variations are closely related 
to variations of the ionizing continuum ($\lambda < 912$\,\AA,
$h\nu \geq 13.6\,{\rm  eV}$).
\end{enumerate}

This information provides the crucial underpinning for the assumptions
detailed below. But before we proceed any further, we need to 
remind ourselves of some of the basic characteristics of the
line-emitting gas, much of which we derive by comparing the
observed line spectra with the predictions of photoionization
equilibrium computer codes (see the contribution by Netzer 
for a more complete discussion). In general, 
photoionization equilibrium models of the line-emitting clouds are 
parameterized by the shape of the ionizing continuum,
elemental abundances, and an ``ionization parameter'' 
\begin{equation}
\label{eq:Udef}
U = \frac{Q({\rm H})}{4\pi r^2 c n_{\rm e}} \mbox{\ \ ,}
\end{equation}
where 
\begin{equation}
Q({\rm H}) = \int^{\infty}_{\nu_1} \frac{L_{\nu}}{h\nu} \ d\nu
\end{equation}
is the number of hydrogen-ionizing ($h\nu_1 = 13.6$\,eV)
photons emitted per second by the central source. 

Essentially,
$U$ characterizes the ionization balance within the
cloud, as $Q({\rm H})/r^2$ is proportional to the
number of photoionizations occurring per second at the
incident face of the cloud, and $n_{\rm e}$ is
proportional to the recombination rate.
For a typical Seyfert 1 galaxy (NGC 5548)
\begin{equation}
Q({\rm H}) \approx 10^{54}\,\ho^{-2}\ {\rm photons\ s}^{-1} \mbox{\ \ ,}
\end{equation}
where $\ho$ is the Hubble constant in units of 100\,\Hubble.
By taking the electron density to be $n_{\rm e} \approx
10^{11}$\,\cc and the distance to the central source to be
10 light days, we find $U\approx 0.1$.
The line-emitting clouds are ionized to the 
``Str\"{o}mgren depth'',
\begin{equation}
\label{eq:RS}
R_{\rm S} = \frac{Uc}{\alpha_{\rm B} n_{\rm e}} \approx 10^{11}\ {\rm cm}
\mbox{\ \ ,}
\end{equation}
at which point virtually all of the incident ionizing
photons have been absorbed. In BLR clouds,
the region beyond the Str\"{o}mgren zone is 
partially ionized, $n({\rm H}^+) \approx 0.1n({\rm H}^0)$,
and heated primarily by far infrared radiation\cite{Fe_ea92}
and partly by high-energy X-rays\cite{KwKr81}.
In Eq.~(\ref{eq:RS}), 
$\alpha_{\rm B}$ is the Menzel--Baker case B
recombination coefficient\cite{Os89}.

\subsection{Reverberation Mapping Assumptions}
There are a number of simplifying assumptions that we can
now make:
\begin{enumerate}
\item  The continuum originates in a single central source.
The size of an accretion disk around a supermassive 
(say, $10^{7-8}$\,\Msun) black hole is of the order of   
$10^{13-14}$\,cm. A typical size for the BLR in the
same system would be of order a few light days,
i.e., $\sim10^{16}$\,cm. Note that it is specifically
not required that the continuum source emits radiation
isotropically, though this is a useful starting point.
The ``point-source'' assumption greatly simplifies 
the reverberation process. However, we should mention
that the point-source assumption is probably not applicable to 
X-ray reverberation, as the Fe K$\alpha$ emission and
the hard X-rays that drive this line probably arise in
regions that are of similar size, and perhaps co-spatial\cite{Re00}.
\item Light-travel time is the most important time scale.
We assume specifically that emission-line clouds respond
instantaneously to changes in the continuum flux. The time scale
to re-establish photoionization equilibrium is the 
recombination time, 
\begin{equation}
\label{eq:rectime}
\tau_{\rm rec} = \left(n_{\rm e} \alpha_{\rm B} \right)^{-1}
\approx 40\left(\frac{n_{\rm e}}{10^{11}\,{\cc}}\right)^{-1} {\rm s.}
\end{equation}
The time it takes a Lyman $\alpha$ photon to diffuse outward from
the Str\"{o}mgren depth is about 20 times the direct light-travel time\cite{HuKu80},
\begin{equation}
\tau_{\rm diff} = 20\frac{R_{\rm S}}{c} \approx 
20 \frac{U}{n_{\rm e} \alpha_{\rm B}} \approx 
60\left(\frac{n_{\rm e}}{10^{11}\,{\cc}}\right)^{-1}\ {\rm s.}
\end{equation}
We also need to carry out our reverberation-mapping experiment
on a time scale short enough that the structure of the BLR
can be assumed to be stable. The dynamical, or cloud-crossing, time scale for the BLR is typically
\begin{equation}
\label{eq:dynamical}
\tau_{\rm dyn} =\frac{r}{\vFWHM} \approx 3 - 5 \ {\rm years},
\end{equation}
where \vFWHM\ is the Doppler width of the broad line for which
the response time $\tau=r/c$ has been measured.
Any reverberation-mapping experiment has to be short relative
to the crossing time or the structural information might be
washed out by cloud motions.
\item There is a simple, though not necessarily linear, relationship 
between the observed continuum and the ionizing continuum.
\end{enumerate}

\subsection{The Transfer Equation}

Under these assumptions, the relationship between the continuum and 
emission lines can be written in terms of the ``transfer equation'',
\begin{equation}
\label{eq:TF}
L(V_z,t) = \int^{\infty}_{-\infty} \Psi(V_z,\tau)\, C(t-\tau)\, d\tau 
\mbox{\ \ .}
\end{equation}
Here $C(t)$ is the continuum light curve, 
$L(V_z,t)$ is the emission-line flux at 
line-of-sight velocity $V_z$ and time $t$,
and $\Psi(V_z,\tau)$ is
the ``transfer function'' at $V_z$ and time delay $\tau$.
Inspection of the transfer equation
shows that the transfer function is simply the time-smeared emission-line 
response to a  $\delta$-function outburst in the continuum. Note that
causality requires that the lower limit on the integral is 
$\tau=0$.

Solution of the transfer equation to obtain the transfer function
is a classical inversion problem in theoretical 
physics: the transfer function is essentially 
the Green's function for the system. Unfortunately, 
it is difficult to find a stable solution to such equations,
especially when the data are noisy and sparse, as they usually
are in astronomical applications.

In practice, most analyses  have concentrated on solving the 
velocity-in\-de\-pen\-dent (or 1-d) transfer equation,
\begin{equation}
\label{eq:1dtf}
L(t) = \int^{\infty}_{-\infty} \Psi(\tau)\, C(t-\tau)\, d\tau \mbox{\ \ ,}
\end{equation}
where both $\Psi(\tau)$ and $L(t)$ represent integrals over the 
emission-line width.

The transfer equation is a linear equation. In reality, however, 
the relationship between the observed continuum and the emission-line
response is likely to be nonlinear. We therefore approximate
$C(t) = \bar{C} + \Delta C$ and 
$L(t) = \bar{L} + \Delta L$, where $\bar{C}$ and $\bar{L}$ 
represent constants, usually the mean value of the 
continuum and line flux, respectively. We can then treat
deviations from the mean as linear perturbations and 
the equation we actually solve is
\begin{equation}
\label{eq:deltaTF}
\Delta L(t) = \int^{\infty}_{-\infty} \Psi(\tau)\, \Delta C(t-\tau)\, d\tau
\mbox{\ \ .}
\end{equation}

Most existing data sets are inadequate for transfer-function solution,
and simpler analyses are used. The most commonly used tool in
analysis of AGN variability is cross-correlation, which
will be discussed in detail in Sec.~{\ref{sect:crossc}}.

\subsection{Isodelay Surfaces}
\begin{figure}
\begin{center}
    \leavevmode
  \centerline{\epsfig{file=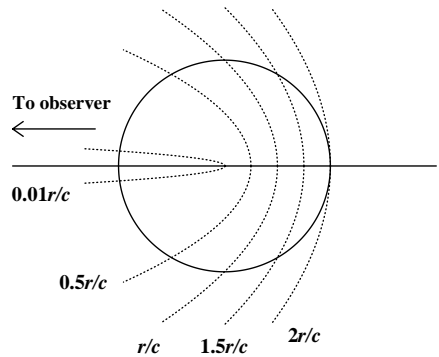,width=6cm,angle=0}}
  \end{center}
\caption{The circle represents a cross-section of a shell
containing emission-line clouds. The continuum source is a point
at the center of the shell. Following a continuum outburst,
at any given time the observer far to the left sees the response of
clouds along a surface of constant time delay, or
isodelay surface. Here we show five isodelay surfaces,
each one labeled with the time delay (in units of the 
shell radius $r$) we would observe relative to the continuum
source. Points along the line of sight to the observer are
seen to respond with zero time delay. The farthest point on
the shell responds with a time delay $2r/c$.
\label{fig:isodelay}}
\end{figure}

Suppose for the moment that the BLR consists of clouds
in a thin spherical shell of radius $r$. Further 
suppose that the continuum light curve is a simple 
$\delta$-function outburst. Continuum photons stream
radially outward and after travel time $r/c$, about 10\%
of these photons (using a typical ``covering factor'')
are intercepted by BLR clouds and are reprocessed into
emission-line photons. An observer at the central source will
see the emission-line response from the entire BLR at a single instant
with a time delay of $2r/c$ following the continuum
outburst. At any other location, however, the summed light-travel
time from central source to line-emitting cloud to observer
will be different for each part of the BLR. 
In the case of a $\delta$-function outburst, at any given
instant, the parts of the BLR that the observer will see
responding are all those for which this total path 
length is identical; at any given time delay, the part of
the BLR that the observer sees responding is the 
intersection of the BLR distribution and an ``isodelay
surface.'' Astronomers, on account of their familiarity
with conic sections,  can readily recognize
that the shape of the isodelay surface is an ellipsoid
with the continuum source at one focus and the
observer at the other; the light-travel time from
central source to BLR cloud to observer is constant for
all points on the ellipsoid. Since the observer is virtually
infinitely distant from the source, the isodelay surface
becomes a paraboloid, as shown schematically in Fig.~\ref{fig:isodelay}.
The figure shows the BLR as a ring intersected by several
isodelay surfaces, labeled in terms of their time delay
in units of $r/c$. Relative to the continuum, points along
the line of sight to the observer are not time delayed
(i.e., $\tau = 0$). Points on the far side of the BLR
are delayed by as much as $2r/c$, the time it takes
continuum photons to reach the BLR plus the time it
takes line photons emitted towards the observer to 
return to the central source on their way to the observer.

\begin{figure}
\begin{center}
    \leavevmode
  \centerline{\epsfig{file=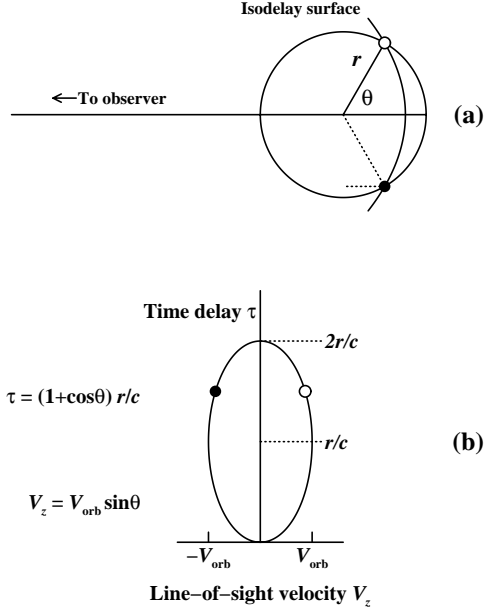,width=7.75cm,angle=0}}
  \end{center}
\caption{The upper diagram shows a ring (or cross-section of
a thin shell) that contains line-emitting clouds,
as in Fig.~\ref{fig:isodelay}. An isodelay surface for
an arbitrary time is given; the intersection of this surface
and the ring shows the clouds that are observed to be responding at this
particular time. The dotted line shows the additional light-travel
time, relative to light from the continuum source, that 
signals reprocessed by the cloud into emission-line photons
will incur (Eq.~(\ref{eq:isod})). In the lower diagram, we
project the ring of clouds onto the line-of-sight velocity/time-delay
($V_z, \tau$) plane,
assuming that the emission-line clouds in the upper diagram
are orbiting in a clockwise direction (so that the cloud
represented by a filled circle is blueshifted and the 
cloud represented by an open circle is redshifted).
\label{fig:tsgeom}}
\end{figure}

Essentially, the transfer function measures the 
amount of line emission emitted at a given Doppler shift 
in the direction of the observer as a function of time delay
$\tau$. The value of the transfer function at time delay $\tau$
is computed by summing the emission in the direction of the
observer at the intersection
of the BLR and the appropriate isodelay surface.
For a thin spherical shell, the intersection of the BLR and
an isodelay surface is a ring of radius $r\sin \theta$,
where the polar angle $\theta$ is measured from the observer's line of
sight to the central source, as shown in Fig.~\ref{fig:tsgeom}. 
The time delay for a particular isodelay surface
is the equation for an ellipse in polar coordinates,
\begin{equation}
\label{eq:isod}
\tau = (1 + \cos\theta)r/c \mbox{\ \ ,}
\end{equation}
as is obvious from 
inspection of Fig.~\ref{fig:tsgeom}. The surface area of the ring 
of radius $r\sin \theta$ and angular width $r\,d\theta$ is
$2\pi (r\sin\theta)\,r\,d\theta$, and assuming that the line response per
unit area on the spherical BLR has a constant value $\varepsilon_0 $,
the response of the ring can be written as
\begin{equation}
\Psi(\theta)=2\pi\varepsilon_0  r^2 \sin\theta\,d\theta \mbox{\ \ ,}
\end{equation}
where $0 \leq \theta \leq 2\pi$.
From Eq.~(\ref{eq:isod}), we can write
\begin{equation}
d\tau = -(r/c)\sin\theta\,d\theta \mbox{\ \ ,}
\end{equation}
so putting the response in terms of $\tau$ rather than 
$\theta$, we obtain
\begin{equation}
\label{eq:tfthinshell}
\Psi(\tau)\,d\tau = \Psi(\theta) \left| \frac{d\theta}{d\tau} \right|
\,d\tau = 2\pi \varepsilon  r c\ d\tau
\end{equation}
for values from $\tau = 0$ ($\theta = 2\pi$) to 
$\tau = 2r/c$ ($\theta = 0$). The transfer function
for a thin spherical shell is thus constant over the
range $0 \leq \tau \leq 2r/c$.

\subsection{Transfer Functions for a Variety of Simple Models}

The simple analytic calculation in the last section was intended to
be illustrative and serve as a reference point. We will
now expand on this with more general geometries, and  
also incorporate information from Doppler motion along the 
line of sight. We will derive some transfer functions for 
other simple models, focusing on three: 
(a) systems of clouds in  circular Keplerian orbits, 
illuminated by an isotropic continuum,
(b) biconical outflows, and 
(c) disks of random inclination.
All these are physically plausible, and can produce ``double-peaked'' 
emission-line profiles, which are sometimes seen in AGNs, 
though not all of these models
necessarily do this. We will start with the
simplest models and progress to more complicated models.
Much of this discussion is drawn from discussions
of transfer functions in the 
literature\cite{WeHo91}$^,$\cite{PRD92}$^,$\cite{Fe_ea92}$^,$\cite{GOG94}$^,$\cite{GoWa96}.

Suppose line-emitting clouds are on a circular orbit 
at inclination $i =90$\deg; imagine that the circle
in Fig.~\ref{fig:tsgeom}a represents this orbit seen face on. The
line response from the
clouds at the intersection of an arbitrary isodelay surface
and the circular orbit 
will be at time delay $\tau = (1 + \cos \theta)r/c$ and
line-of-sight velocities $V_z = \pm V_{\rm orb}\sin\theta$,
where $V_{\rm orb} = (GM/r)^{1/2}$, the circular orbital speed.
It is easy to see that the circular orbit projects to 
an ellipse in the line-of-sight velocity/time-delay 
($V_z, \tau$) plane with semiaxes $V_{\rm orb}$ and $r/c$, 
as shown in Fig.~\ref{fig:tsgeom}b. This simple example is
important because it is straightforward to generalize
it to both disks (rings of different radii) and
shells (rings at different inclinations).

\begin{figure}
\begin{center}
    \leavevmode
  \centerline{\epsfig{file=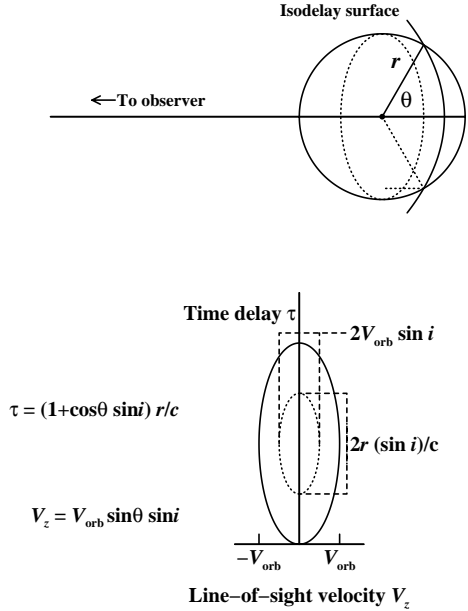,width=7.75cm,angle=0}}
  \end{center}
\caption{Here we take the circular ring shown in Fig.~\ref{fig:tsgeom} and show how its projection on 
$V_z, \tau$ changes as the inclination is decreased from
90\deg; in the time-delay direction, the axis contracts
to $2r \sin i /c$, and in the velocity direction it
decreases to $2V_{\rm orb} \sin i$.
\label{fig:tsgeom2}}
\end{figure}

\begin{figure}
\begin{center}
    \leavevmode
\vspace{4cm}
  \end{center}
\caption{Transfer function for a thin spherical shell.
The upper left panel shows in grey scale the two-dimensional transfer
function, i.e., the observed emission-line response as
a function of line-of-sight velocity $V_z$ and time
delay $\tau$. The upper right panel shows the one-dimensional
transfer function, i.e., the two-dimensional transfer function
integrated over $V_z$, which is the response of the
total emission line as a function of time
(cf.\ Eq.~(\ref{eq:tfthinshell})). The lower left
panel shows the emission-line response integrated over time
delay; this is the profile of the variable part
of the line. The specific model here is comprised of
emission-line clouds in circular orbits of radius 
$r = 10$\,light days and random inclinations 
($0\deg \leq i \leq 180\deg$), orbiting
around a central black hole of mass $10^8$\,\Msun. In
this model, the line emission from each cloud is isotropic
($A=0$).
\label{fig:tstfa0}}
\end{figure}

First we consider the generalization to a shell. We can construct
a shell from a distribution of circular orbits, with inclinations
ranging from $i = 0\deg$ to $i = 180$\deg. As we decrease the
inclination of the circular orbit in Fig.~\ref{fig:tsgeom}a from 
$i = 90\deg$, we see that the range of time delays will
decrease from [0, $2r/c$] to [$(1-\sin i)r/c$, $(1+\sin i)r/c$],
and similarly the line-of-sight velocity range will decrease from
[$-V_{\rm orb}$, $+V_{\rm orb}$] to 
[$-V_{\rm orb}\sin i$, $+V_{\rm orb}\sin i$],
as we show schematically in Fig.~\ref{fig:tsgeom2}.
At the limiting
case $i=0\deg$, the time delays all contract to $r/c$, since
the light travel-time paths for all points on a face-on ring are
the same, and the velocities all contract to 
$V_z = 0$, because the orbital velocities are now
perpendicular to the line of sight. Thus, the transfer
function looks like a series of ellipses as in Fig.~\ref{fig:tsgeom2}b
that with decreasing inclination
contract down to a single point at $V_z =0$ and $\tau=r/c$
when $i=0\deg$.
We can construct such a transfer function
by using a Monte Carlo method that places BLR clouds 
randomly across the surface of the shell, and the result
we get is shown in Fig.~\ref{fig:tstfa0}, 
which corresponds to a thin shell of 
radius 10 light days and a central mass of $10^8\,\Msun$. We 
have also integrated this transfer function over 
$\tau$ and $V_z$ to obtain the emission-line
profile and the one-dimensional transfer function, respectively.
In the particular case of a thin spherical shell, we see
that both of these are simple rectangular functions,
as we showed analytically for the one-dimensional transfer
function (Eq.~(\ref{eq:tfthinshell})).

The first complication that we might consider is
anisotropic line emission by the BLR clouds. Physically,
this will occur if the BLR clouds themselves are optically
thick in the lines as well as the continuum. In this case,
most of the line radiation emitted by the clouds will be
from the side of the cloud facing the continuum source,
i.e., the line emission is directed back towards the continuum
source. A simple parameterization of asymmetric line emission
is to describe the apparent emissivity of a cloud as
\begin{equation}
\varepsilon(\theta) = \varepsilon_{0} 
\left( 1 + A\cos\theta \right) \mbox{\ \ ,}
\end{equation}
where $\varepsilon_{0}$ is constant and the
parameter $A=0$ for isotropic emission
and $A = 1$ for completely anisotropic emission; the latter
case is appropriate for spherical clouds 
with inward-facing surfaces that are uniformly bright.
In principle, $A$ can be estimated by photoionization
modeling, though in practice the values are highly
uncertain on account of limitations in the accuracy
of the radiative transfer codes (see the contribution by Netzer).
It is certainly expected that $A\approx 1$ for \Lya,
at least, and models suggest approximate values
$A\approx 0.7$ for \civ\ and
$A\approx 0.8$ for \Hbeta\cite{Fe_ea92}.
The main effect of anisotropic emission is to increase
the measured lag for a given geometry because the
apparent response of the near side of the BLR is
suppressed. For a thin shell of radius $r$, the
mean time delay is $\tau = (1 + A/3)r/c$. Fig.~\ref{fig:tstfa1}
shows the transfer function for the same thin spherical
shell model of Fig.~\ref{fig:tstfa0}, but in this case with highly
anisotropic ($A = 1$) line response.

\begin{figure}
\begin{center}
    \leavevmode
\vspace{4cm}
  \end{center}
\caption{Transfer function for a thin spherical shell,
as in Fig.~\ref{fig:tstfa0}, except with completely anisotropic ($A=1$) line
emission, i.e., all of the emission line flux from each
cloud is directed back towards the continuum source.
\label{fig:tstfa1}}
\end{figure}

\begin{figure}
\begin{center}
    \leavevmode
  \centerline{\epsfig{file=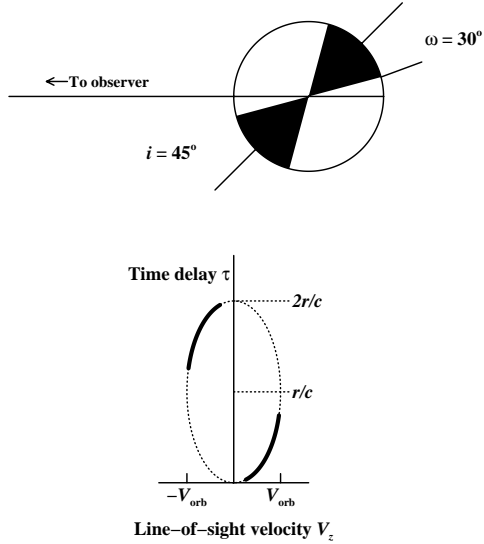,width=7cm,angle=0}}
  \end{center}
\caption{A ring-like line-emitting region and its projection
into the $V_z, \tau$ plane is shown, as in 
Figs.~\ref{fig:tsgeom} and \ref{fig:tsgeom2}, though in
this case with a biconical continuum geometry. In the case shown
here, the beam semi-opening angle is $\omega = 30$\deg,
and the beam is inclined to the observer by
$i = 45$\deg. Again assuming that clouds on the ring
are orbiting in the clockwise direction, we highlight in
the lower diagram the regions in the ($V_z, \tau$) plane
that will produce an emission-line response;
there is no emission-line response from clouds not
illuminated by the continuum beam.
\label{fig:bcgeom}}
\end{figure}

In addition to anisotropic line response, we can
also consider anisotropic illumination of the BLR
clouds by the continuum source.
As an example, consider the case where BLR 
clouds are illuminated by biconical beams of
semi-opening angle $\omega$; the limiting case as
$\omega$ approaches zero would be a narrow jet-like
pencil beam, and the case $\omega = 90\deg$ corresponds
to an isotropic continuum. We 
start by examining the response of an edge-on 
($i = 0\deg$) circular ring, as we show in Fig.~\ref{fig:bcgeom},
which is exactly like Fig.~\ref{fig:tsgeom}, but with only
part of each orbit illuminated by the continuum source.
We earlier generalized the result for a shell,
going from Fig.~\ref{fig:tsgeom} to Fig.~\ref{fig:tsgeom2}
as we decreased the inclination of the ring; doing this again, we see
in Fig.~\ref{fig:tstfbc} how the two-dimensional transfer function
is altered by anisotropic illumination. We note specifically
the absence of any response near $\tau = 0$ and
$\tau = 2r/c$ since the bicones shown do not illuminate
BLR clouds directly along the observer's axis. Similarly there is
also no response around $\tau = r/c$ due to the absence
of response of clouds at $\theta\approx 90\deg$. 

\begin{figure}
\begin{center}
    \leavevmode
  \centerline{\epsfig{file=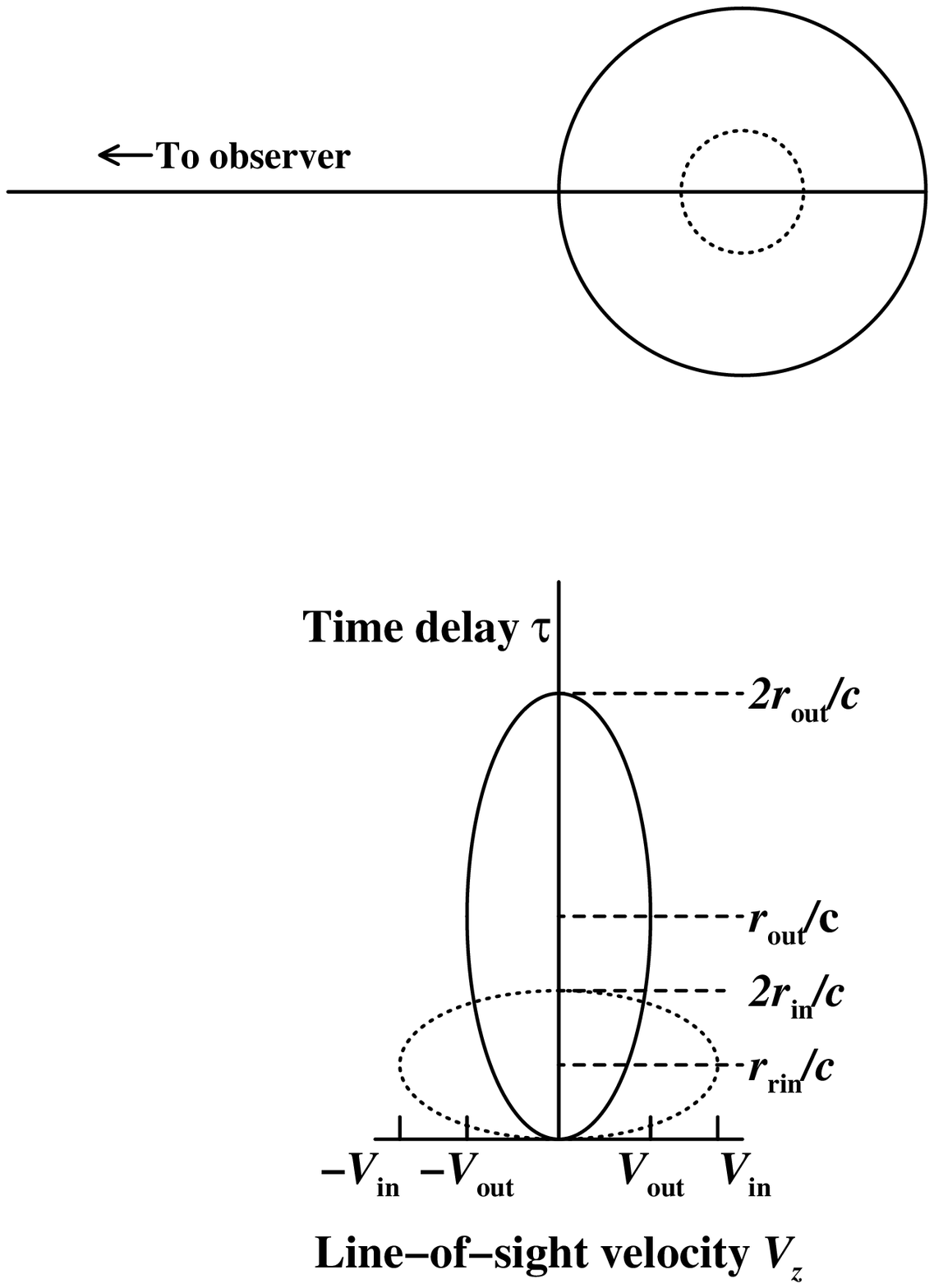,width=6.5cm,angle=0}}
  \end{center}
\caption{Two ring-like regions (as in Fig.~\ref{fig:tsgeom}),
and their projected response on the ($V_z, \tau$) plane.
\label{fig:twoshell}}
\end{figure}

So far we have restricted our attention to ``thin''
geometries, i.e., single orbits and shells. Generalization
to ``thick'' geometries, e.g., disks and shells with different
inner and outer radii, is trivial: the response of a disk
can be computed by integrating over a series of circular orbits,
and the response of a thick shell is obtained by integrating
over a series of shells. In Fig.~\ref{fig:twoshell}, we illustrate this
concept by showing the response from two rings.
\begin{figure}
\begin{center}
    \leavevmode
\vspace{4cm}
  \end{center}
\caption{Transfer function for a thin spherical shell,
as in Fig.~\ref{fig:tstfa0}, except with an anisotropic
continuum source. In the example shown, the beam
opening angle is $\omega = 30$\deg\ and the
inclination is $i = 45$\deg, as in
Fig.~\ref{fig:bcgeom}.
\label{fig:tstfbc}}
\end{figure}
Thus far, the free parameters we
have dealt with are the radius $r$, the line anisotropy
factor $A$, and in the case of non-spherically symmetric
systems, the inclination $i$ and, in the case of
biconical illumination, the beam half-angle $\omega$.
Thick geometries now introduce inner and outer radii,
$r_{\rm in}$ and $r_{\rm out}$, respectively, and
a distance-dependent responsivity per unit volume
(or per unit area, for a disk), which we can parameterize as
$\varepsilon(r) \propto \varepsilon_{0} r^{\beta}$. The
index $\beta$ allows us to condense 
model-dependence into a single parameter that 
will account for effects due to 
geometrical $r^{-2}$ dilution of the continuum, a
distance-dependent covering factor, etc.
In Figs.~\ref{fig:tktfb00} and \ref{fig:tktfbm4}  
we show transfer functions
for thick spherical shell systems with $A=0$ and identical
values of $r_{\rm in}$ and $r_{\rm out}$, but differing
radial-response indices $\beta$; the effect of increasing
$\beta$ is to enhance the relative response of material at
larger radii; the limiting cases where 
$\beta \rightarrow -\infty$ and
$\beta \rightarrow +\infty$ correspond to the response functions
of thin shells of radius $r_{\rm in}$ and
$r_{\rm out}$, respectively. We will show additional examples
below.

We consider now a thick shell system that is illuminated
by an anisotropic continuum source. Again, we assume that the
line-emitting clouds are in circular Keplerian orbits of
random inclination. We show examples that are identical except
for the continuum beam width and inclination in
Figs.~\ref{fig:tkbicone1} and \ref{fig:tkbicone2}. An
important thing to notice is that in one case (Fig.~\ref{fig:tkbicone1}) the
observer is outside of the continuum beam (i.e., $i > \omega$) and
in the other case (Fig.~\ref{fig:tkbicone2}) the
observer is inside of the continuum beam (i.e., $i < \omega$);
when the observer is {\em inside} the beam, the line profile
is {\em single-peaked}, and when the observer is {\em outside}
the beam, the line profile is {\em double-peaked}.

\begin{figure}[t]
\begin{center}
    \leavevmode
\vspace{4cm}
  \end{center}
\caption{Transfer function for a thick spherical shell
with $r_{\rm in} = 2$ lt-days, $r_{\rm out} = 10$ lt-days, 
and radial responsivity index $\beta = 0$.
\label{fig:tktfb00}}
\end{figure}
\begin{figure}
\begin{center}
    \leavevmode
\vspace{4cm}
  \end{center}
\caption{Transfer function for a thick spherical shell
exactly as described in Fig.~\ref{fig:tktfb00}, except with 
radial responsivity index $\beta = -4$.
\label{fig:tktfbm4}}
\end{figure}

\begin{figure}[t]
\begin{center}
    \leavevmode
\vspace{4cm}
  \end{center}
\caption{Transfer function for a thick spherical shell
with $r_{\rm in} = 2$ lt-days, $r_{\rm out} = 10$ lt-days, 
radial responsivity index $\beta = -2$, and isotropic
line response $A=0$.
In this model, the shell is illuminated by a biconical
continuum with semi-opening angle $\omega = 30\deg$
and inclination $i=45\deg$, i.e., the observer is
outside the continuum beam. Note the double-peaked line
profile, and contrast this with the model shown in
Fig.~\ref{fig:tkbicone2}.
\label{fig:tkbicone1}}
\end{figure}
\begin{figure}
\begin{center}
    \leavevmode
\vspace{4cm}
  \end{center}
\caption{Transfer function for a thick spherical shell
exactly as described in Fig.~\ref{fig:tkbicone1},
except that in this model the shell is illuminated by a biconical
continuum with semi-opening angle $\omega = 75\deg$
and inclination $i=15\deg$, i.e., the observer is
well inside the continuum beam. Note the single-peaked line
profile, and contrast this with the model shown in
Fig.~\ref{fig:tkbicone1}.
\label{fig:tkbicone2}}
\end{figure}

Another simple kinematic model for the BLR consists of clouds
in spherical outflow. The transfer functions for such cases
are quite distinctive; an example of a two-dimensional tranfer
function for a kinematic field with radial velocity 
$V_r \propto r$ for $r$ less than some maximum distance
$r_{\rm out}$ is shown in Fig.~\ref{fig:oftfiso}.
This velocity field
corresponds to either a ballistic outflow
or a flow undergoing constant acceleration. At
$\tau =0$ we see response from all the material along the
line of sight to the continuum source, which ranges
from $V_z =0$ for the material closest to the central source
to $V_z = -V(r_{\rm out})$ for the gas farthest from the continuum
source. As $\tau$ increases, we begin to see response from the
far side of the BLR where the line of sight velocities $V_z$
become positive. The range of line--of-sight velocities
decreases as the isodelay surfaces get farther from the
line of sight. At $\tau=r_{\rm out}/2$, the isodelay surface
no longer crosses any clouds moving towards the observer,
and the gas moving fastest away from the observer is that
along the line of sight ($\theta = 0$) with 
$V_z = V(r_{\rm out})/2$. At $\tau=2r_{\rm out}/c$,
only the response from the antipodal point is seen, and the
transfer function is contracted to the single point
at $[2r_{\rm out}/c, V(r_{\rm out})]$.

The case of biconical outflows (which are possibly relevant,
as they are certainly seen in the NLR and might well apply
to at least a component of the BLR) can be 
dealt with by restricting the response to certain values
of $\theta$; an example is shown in Fig.~\ref{fig:oftf45}.

\begin{figure}
\begin{center}
    \leavevmode
  \centerline{\epsfig{file=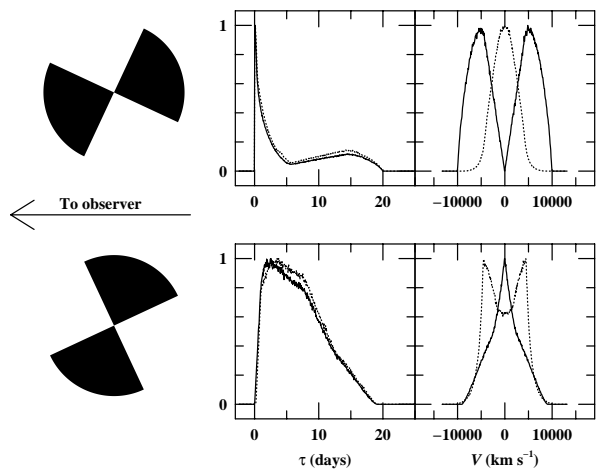,width=8cm,angle=0}}
  \end{center}
\caption{Response of two biconical models with different
kinematics. The left hand column shows the two geometries
considered, both with beam half-opening angles 
$\omega = 40$\deg; the top row is at inclination
$i=25$\deg\ (i.e., the observer is in the beam) and
the bottom row is $i=65$\deg\ (i.e., the observer is out
of the beam). One-dimensional transfer functions
(middle column) and line profiles (right column) are
shown. The solid line is a biconical outflow model,
and the dashed line represents circular Keplerian
orbits of random inclination. The one-dimensional
transfer functions are identical, but the profiles
are greatly different, as described in the text.
\label{fig:comp1}}
\end{figure}

We have now seen that both orbital and outflow models
can produce similar emission-line profiles; if the
line-emitting gas is confined to a bicone, either
because of the gas distribution or the ionizing-photon
distribution, one can get a single-peaked or double-peaked
line profile. The two situations can be easily distinguished,
however, by their two-dimensional transfer functions
(or equivalently, the combination of their one-dimensional
transfer functions and their line profiles).
In Fig.~\ref{fig:comp1}, we directly compare the one-dimensional
transfer functions and line profiles for two thick-shell
models: (1) emission-line clouds in a biconical outflow
and (2) clouds in circular Keplerian orbits of
random inclination,  illuminated by a biconical
continuum source. In both cases, the beams (one
radiation, one matter) have the same half-opening
angle ($\omega=40$\deg) and two different inclinations
are shown; $i = 25$\deg\ in the top row
(i.e. the observer is in the beam, as indicated in the left-hand
column illustrations of the geometry), and 
$i = 65$\deg\ in the bottom row (i.e., the observer is out
of the beam). The distribution of line-emitting clouds is the
same, regardless of how the clouds are moving, so in these
two cases the one-dimensional transfer functions ought to be
very similar, which is indeed the case, as seen in the middle column of 
Fig.~\ref{fig:comp1}. The right-hand column shows the
line profiles, which are however very different. Consider the
case $i=25$\deg\ (top row): in the case of outflow, the
line-emitting material in the beam is moving radially
outward, giving relatively highly 
blueshifted (near side) and redshifted (far side) emission,
but little emission near $V_z = 0$, since there is no
line-emitting material moving transverse to the line of
sight. In the case of clouds in circular orbits illuminated
by an anisotropic beam, the cloud motions are perpendicular
to their radial vectors, so most of the line-emitting material
is at low Doppler shift as the gas motions through
the beams are predominantly transverse. Now consider the
higher-inclination case ($i=65$\deg; bottom row): in the
case of radial outflow, the gas motions in this case are now
primarily transverse so the Doppler shifts are smaller.
However, in the case of circular orbits, the material in
the beam is moving primarily along the line of sight, and
there is a deficiency of material at $V_z \approx 0$.
\begin{figure}[t]
\begin{center}
    \leavevmode
\vspace{4cm}
  \end{center}
\caption{Transfer function for a spherical outflow,
with outflow velocity $V \propto r$. For this model,
$r_{\rm in} = 0.1$ lt-days, $r_{\rm out} = 10$ lt-days, 
$V(r_{\rm out}) = 10000$\,\kms.
\label{fig:oftfiso}}
\end{figure}
\begin{figure}
\begin{center}
    \leavevmode
\vspace{4cm}
  \end{center}
\caption{Transfer function for a biconical outflow,
with parameters as in Fig.~\ref{fig:oftfiso} except
that the outflowing gas is confined to a bicone
of half-width $\omega = 30$\deg\ at inclination
$i = 45$\deg.
\label{fig:oftf45}}
\end{figure}
Note that either kinematic field can give either 
double-peaked or single-peaked profiles: a simple 
generalization is that double-peaked profiles are
produced in outflow models when the observer's line-of-sight
is {\em in the beam} (i.e., $i \ltsim \omega$)
and in orbital models when the line of sight 
is {\em out of the beam} ($i \gtsim \omega$). Neither the profiles
nor the one-dimensional transfer functions individually tell us
much about the BLR kinematics and velocity field,
but together they can tell us a lot.
Information on both $\tau$ and $V_z$, i.e.,
the two-dimensional transfer function, greatly reduces the
ambiguities.

Finally, for completeness, we consider the case of
an inclined disk, as this is the classic geometry
for producing a double-peaked line profile. The response
of a disk-like BLR can be computed by integrating the
response of rings of different radii. In Fig.~\ref{fig:disk}, we show the transfer function and line profile
for a disk at intermediate inclination ($i=45$\deg).
Identical line profiles can be obtained for different
inclinations simply by suitably adjusting the central
mass, but the transfer function allows us to reduce 
the ambiguity between possible disk models.

\begin{figure}
\begin{center}
    \leavevmode
\vspace{4cm}
  \end{center}
\caption{Transfer function for a thin disk
at intermediate inclination ($i=45$\deg).
Note that the line profile is double peaked, and the
one-dimensional transfer function is similar to those
for spherical shells.
\label{fig:disk}}
\end{figure}

\subsection{Transfer Function Recovery}
With data of sufficient quality, transfer 
functions for different plausible models are distinguishable from 
one another. This is of course, why we place so much emphasis
on them: if we can determine the transfer function for
a particular line, we have very strong constraints on the
geometry and kinematics of the responding region, and if the
BLR has a spherically or azimuthally symmetric structure, it
may be possible to determine the BLR kinematics and
structure with fairly high accuracy and strongly constrain
the emission-line reprocessing physics. The operational goal
of reverberation mapping experiments is thus to determine
the transfer functions for various emission lines in AGN 
spectra. 

How can we determine the transfer functions from observational data?
Inspection of the transfer equation (Eq.~(\ref{eq:TF})) immediately
suggests Fourier inversion (i.e., the method of Fourier 
quotients), which was the formulation outlined by Blandford \& McKee\cite{BlMc82}.
 We define the Fourier transform of the continuum light
curve $C(t)$ as
\begin{equation}
C^*(f) = \int_{-\infty}^{+\infty} C(t)\,e^{-i 2\pi f t}\ dt
\end{equation}
and similarly define the Fourier transform of the line light curve.
By the convolution theorem\cite{Br65}, the transfer function becomes
\begin{equation}
L^* = \Psi^* \times C^* \mbox{\ \ ,}
\end{equation}
so that $\Psi^* = L^*/C^*$, and the transfer function is obtained from
\begin{equation}
\Psi(\tau) = \int_{-\infty}^{+\infty} \Psi^*(f)
e^{i 2\pi f t}\ dt \mbox{\ \ .}
\end{equation}

In practice, however, Fourier methods are not used as they
are inadequate when applied to data that are relatively
noisy (i.e., flux uncertainties are only a factor of a few to
several smaller than the variations) and which are
limited in terms of both sampling rate, which is in any case
usually irregular, and duration.
Simpler methods, like cross-correlation (Sec.~{\ref{sect:crossc}}),
can be applied with success, though
with limited results. Cross-correlation, we will see, can
give the first moment of the transfer function, but little else.

In principle, more powerful methods can be used to 
attempt to recover transfer functions. The most commonly used
is the Maximum Entropy Method (MEM)\cite{Ho94}.
MEM is a generalized version of maximum likelihood methods, such as 
least-squares. The difference is that in the method of least 
squares, a parameterized model is fitted to the data, whereas 
MEM finds the ``simplest'' (maximum entropy) solution, balancing model 
simplicity and realism. Examples of MEM solutions will
be shown in Sec.~{\ref{sect:results}}. Other methodologies
that have been employed for transfer-function solution
include subtractively optimized local averages
(SOLA)\cite{PiWa94} and
regularized linear inversion\cite{KrDo95}.

\section{Cross-Correlation Analysis}
\label{sect:crossc}

\subsection{Fundamentals}
Cross-correlation analysis is the tool most commonly used in the analysis
of multiple time series. Because its application to
astronomical time series is often misunderstood and has historically
been rather contentious, it merits special attention.
Important steps in the development of cross-correlation
analysis as applied to AGN variability studies 
can be found in the literature\cite{GaSp86}$^,$\cite{GaPe87}$^,$\cite{EdKr88}$^,$\cite{MaNe89}$^,$\cite{KoGa91}.

\begin{figure}
\begin{center}
    \leavevmode
  \centerline{\epsfig{file=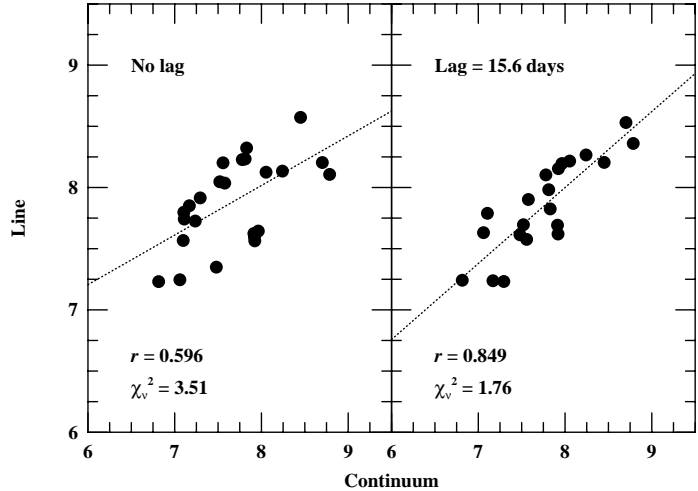,width=9.3cm,angle=0}}
  \end{center}
\caption{The left-hand panel shows simultaneous measurements
of  \Hbeta\ emission-line and optical continuum
fluxes for the Seyfert galaxy Mrk 335$^{69}$.
These consist of 24 measurements made on an approximately weekly
basis over one observing season.  The right-hand panel shows the same
emission-line fluxes paired with continuum values from
15.6 days earlier; i.e., the emission-line fluxes are better
correlated with {\em earlier} rather than {\em current} 
continuum values on account of light travel-time delays.
When the time lag is taken into account, the fit improves.
\label{fig:datacor}}
\end{figure}

Cross-correlation analysis is basically a generalization of standard
linear correlation analysis, which provides us with a good place to 
start. Suppose we obtain repeated spectra of one of the brighter
Seyfert galaxies, and we want to determine whether or not the
variations in the \Hbeta\ emission line and the optical continuum
are correlated (which was an interesting question 20 years ago,
even before emission-line time delays were considered).
The first thing you would do is plot the \Hbeta\ flux against the 
continuum flux, as in the left-hand
panel of Fig.~\ref{fig:datacor}, which shows that the two variables are
indeed correlated. A measure of the strength of the
correlation is given by the {\em correlation coefficent},
\begin{equation}
r = \frac{ 
	\sum_{i=1}^{N}  \left( x_i - \bar{x} \right)
                      \left( y_i - \bar{y} \right)}
{\left( \sqrt{\sum_{i=1}^{N}  \left( x_i - \bar{x} \right)^2 } \right) 
 \left( \sqrt{\sum_{i=1}^{N}  \left( y_i - \bar{y} \right)^2 } \right) }
\mbox{\ \ ,}
\label{eq:rdef}
\end{equation}
where there are $N$ pairs of values $(x_i, y_i)$ and their respective
means are $\bar{x}$ and $\bar{y}$. When the two variables $x$ and
$y$ are perfectly correlated, $r=1$. If they are perfectly anticorrelated,
$r=-1$. If they are completely uncorrelated, $r=0$. For the data shown
in the left panel of 
Fig.~\ref{fig:datacor}, $r=0.596$; for 24 pairs of points, as shown
here, this means that the correlation is significant at the 
$\gtsim99.8$\% confidence level (i.e., the chance that the two variables are
in fact completely uncorrelated and the correlation we find is 
spurious is less than 0.02\%. Confidence levels for linear correlation
can be found in standard statistical tables\cite{Be69})

\begin{figure}
\begin{center}
    \leavevmode
  \centerline{\epsfig{file=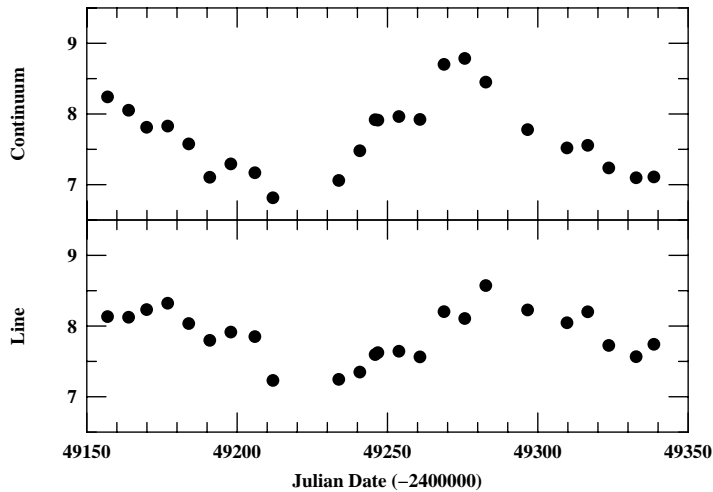,width=9.5cm,angle=0}}
  \end{center}
\caption{The \Hbeta\ emission-line and optical continuum
fluxes for Mrk 335, as shown in Fig.~\ref{fig:datacor},
are plotted as a function of time. It is clear from the
figure that the continuum and emission-line fluxes are
well-correlated, and that the correlation can be improved
by a linear shift in time of one time series relative
to the other. The optimum linear correlation occurs
by shifting the emission-line light curve backwards by
15.6 days.
\label{fig:timecor}}
\end{figure}

While this is quite a good correlation, we see something more remarkable
if we plot both variables as functions of time (i.e., as
{\em light curves}), as seen in Fig.~\ref{fig:timecor}. We see that 
the patterns of variation are very similar, except that the emission-line
light curve is delayed in time, or ``lagged,'' relative to the continuum
light curve. It is obvious that the correlation between the continuum
and emission-line fluxes would be even better if we allowed a linear
shift in time between the two light curves in order to line up
their prominent maxima and minima. This is what cross-correlation does.

The first operational problem in computing a cross-correlation is 
also immediately apparent: since each point in one light curve must
be paired with a point in the other light curve, it is obvious that
the data should be regularly spaced. The cross-correlation is then
evaluated as a function of the spacing between the
interval between data points $\Delta t$ using the pairs
[$x(t_i), y(t_i + N\Delta t)$] for all integers $N$. 
Unfortunately, regularly sampled data are almost never found in Astronomy;
ground-based programs have weather to contend with, and even
satellite-based observations are almost never regularly spaced in time.
The essence of the cross-correlation problem in Astronomy is dealing with 
time series that are not evenly sampled. Moreover, the light
curves are often limited in extent and are noisy.

\begin{figure}
\begin{center}
    \leavevmode
  \centerline{\epsfig{file=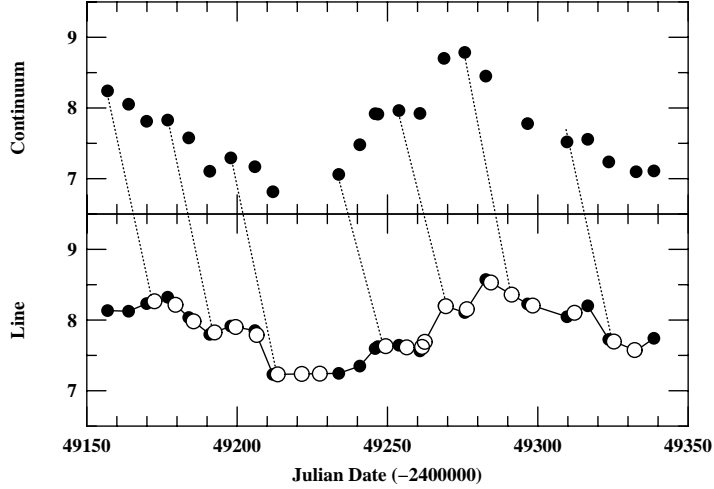,width=9.5cm,angle=0}}
  \end{center}
\caption{Continuum and emission-line light curves for Mrk 335,
as in Fig.~\ref{fig:timecor}. This illustrates the interpolation
method commonly used in cross-correlation. In this figure,
the emission-line light curve is made continuous through
linear interpolation between data points.
Actual continuum observations are then
paired with interpolated emission-line values to compute
the correlation coefficient for a particular time delay.
In this example, we show interpolated emission-line fluxes
that are time-delayed relative to the continuum by
15.6 days, which is the value at which the cross-correlation
function peaks. As a visual aid, dotted lines join a few of the 
data pairs. Notice how the first few points of the emission-line
series and the last few points of the continuum series remain
unused.
\label{fig:interplc}}
\end{figure}

\begin{figure}
\begin{center}
    \leavevmode
  \centerline{\epsfig{file=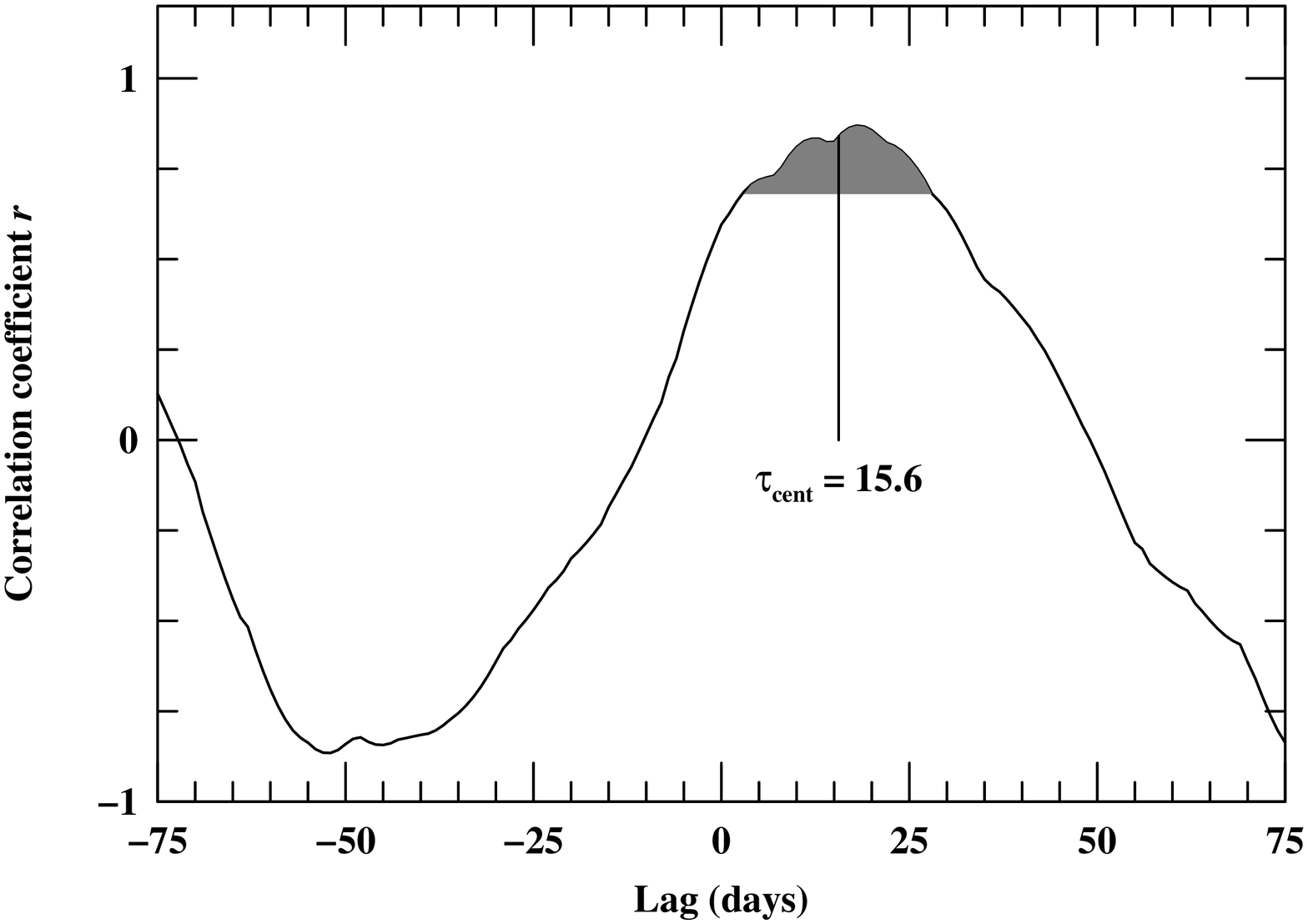,width=9cm,angle=0}}
  \end{center}
\caption{The interpolation cross-correlation function for
the Mrk 335 data shown in the previous figures.
The shaded area indicates points with values
$r \geq 0.8r_{\rm peak}$, which are the points used in computing
the centroid, which is also indicated.
\label{fig:ccf}}
\end{figure}

For well-sampled
series as in Fig.~\ref{fig:timecor}, the sampling problem can be
dealt with in a straightforward fashion. The simple, effective solution
is to interpolate one series between the actual data points, and 
use the interpolated points in the cross-correlation. We illustrate
this schematically in Fig.~\ref{fig:interplc}. We can then
compute the cross-correlation function  CCF($\tau$),
as shown in Fig.~\ref{fig:ccf}, and 
the step size we use for $\tau$ is now somewhat arbitrary.
At each value of the lag $\tau$ we compute 
$r$ as in Eq.~(\ref{eq:rdef}).
For the example we have been using, we find that the
CCF is maximized when points in the continuum light curve are
matched to those in the emission-line light curve with a delay
of 15.6\,days. If we plot the shifted emission-line
values versus the continuum values (as we have done in the
right-hand panel of Fig.~\ref{fig:datacor})
and again perform a linear correlation analysis, we find
that the fit has improved, with $r=0.849$ and
$\chi^2_{\nu} = 1.76$.

\subsection{Relationship to the Transfer Function}

It is useful at this point to examine how the CCF is
related to more fundamental quantities. Defined
as a function of the continuous variable $\tau$, the
formal definition of the cross-correlation function is
\begin{equation}
\label{eq:ccfdef}
{\rm CCF}(\tau) = \int^{\infty}_{-\infty} L(t)
C(t-\tau)\ dt \mbox{\ \ .}
\end{equation}
If we use Eq.~(\ref{eq:1dtf}) to replace $L(t)$ 
in this equation, we obtain
\begin{eqnarray}
{\rm CCF}(\tau) & = & \int^{\infty}_{-\infty} C(t-\tau)
\int^{\infty}_{-\infty} \Psi(\tau')C(t-\tau')\ d\tau'\ dt \nonumber \\
&= & \int^{\infty}_{-\infty} \Psi(\tau')
\int^{\infty}_{-\infty} C(t-\tau') C(t-\tau)\ dt\ d\tau' \mbox{\ \ .}
\end{eqnarray}
Comparing the inner integral with Eq.~(\ref{eq:ccfdef}), we
see that it is the cross-correlation of the continuum light
curve with itself, i.e., the ``autocorrelation function'',
\begin{equation}
{\rm ACF}(\tau) =\int^{\infty}_{-\infty} C(t)
C(t-\tau)\ dt \mbox{\ \ .} 
\end{equation}
Thus, we see that the cross-correlation function is
the convolution of the transfer function with the 
continuum autocorrelation function,
\begin{equation}
{\rm CCF}(\tau) = \int^{\infty}_{-\infty} \Psi(\tau')
{\rm ACF}(\tau-\tau')\ d\tau' \mbox{\ \ .}
\end{equation}

The centroid of the CCF is a usually quoted quantity,
\begin{equation}
\tau_{\rm cent} = \frac{ \int\! \tau\, {\rm CCF}(\tau)\, d\tau}
{\int {\rm CCF}(\tau)\ d\tau} \mbox{\ \ ,}
\end{equation}
which is related to the centroid of the transfer function
\begin{equation}
\tau^*_{\rm cent} = \frac{\int\! \tau\, \Psi(\tau)\, d\tau}
{\int\! \Psi(\tau)\ d\tau} \mbox{\ \ .}
\end{equation}
It is important to note that these two quantities are the same
only in the limit where the light curves extend to infinity.
Operationally, the centroid is computed using only points
around the most significant peak, usually those points for which
${\rm CCF}(\tau) \geq 0.8 r_{\rm peak}$, where 
$r_{\rm peak}$ is the maximum value of the CCF. Sometimes the
location of the peak $\tau_{\rm peak}$ 
(${\rm CCF}(\tau_{\rm peak}) = r_{\rm peak}$) is the statistical
quantity used.

Cross-correlation
does not necessarily give a clean and unambiguous measure of the relationship
between two time series. In particular, the CCF is a convolution
of the transfer function with the continuum ACF, 
so the CCF depends on the continuum behavior. To illustrate
this, we show in Fig.~\ref{fig:placf} examples of 
light curves for various
power-law PDSs (Eq.~(\ref{eq:PDS})) and the very different ACFs computed for
each example. 

\begin{figure}[h]
\begin{center}
    \leavevmode
\vspace{4cm}
  \end{center}
\caption{The left-hand column shows artificial light curves
generated with different power-law indices $\alpha$
for the power-density spectra, $P(f) \propto f^{-\alpha}$.
All light curves consist of 2048 points, normalized by
setting $\Fvar = 0.3$ (Eq.~(\ref{eq:Fvar})). The right-hand column shows the
autocorrelation functions for these light curves.
Cases shown, top to bottom, are:
(a) $\alpha=0$. There is no correlation from one point to
the next, as seen clearly in the ACF, which is essentially
zero except at $\tau = 0$. This is known as ``white noise.''
(b) $\alpha=1$. This is the power spectrum characteristic of
a random walk; each continuum point now has some dependence
on history. This case is sometimes called ``flicker noise''
(see Press$^{78}$ for further interesting discussion of this
case).
(c) $\alpha=1.5$. This appears to be rather typical of
AGN variability.
(d) $\alpha=2$. This power spectrum is characteristic of
a superpositioning of independent events, which is often
referred to as ``shot noise.''
\label{fig:placf}}
\end{figure}

\subsection{Discrete Correlation Methods}

\begin{figure}
\begin{center}
    \leavevmode
  \centerline{\epsfig{file=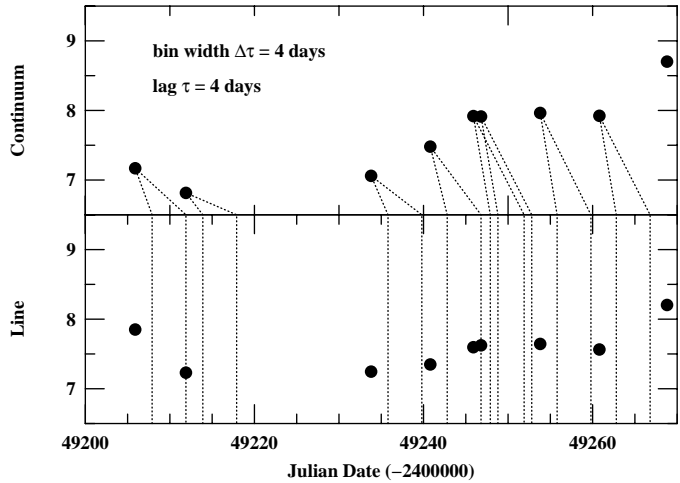,width=9cm,angle=0}}
  \end{center}
\caption{Part of the light curves from the previous figures,
expanded to show DCF bins. The particular example shown is
for a time-lag bin width $\Delta \tau = 4$ days and shows
the location of the bins, pairing real continuum points
with emission-line points within a 4-day window, centered
at a shift of 4 days (i.e., emission-line data points shifted 
by 2--6 days from the continuum points). Note that in this
expanded region there are only three emission-line points
that fall into the bins; in the range shown here, only
two continuum and three emission-line points contribute to 
the computation of the correlation coefficient.
\label{fig:dcfbins}}
\end{figure}

\begin{figure}
\begin{center}
    \leavevmode
  \centerline{\epsfig{file=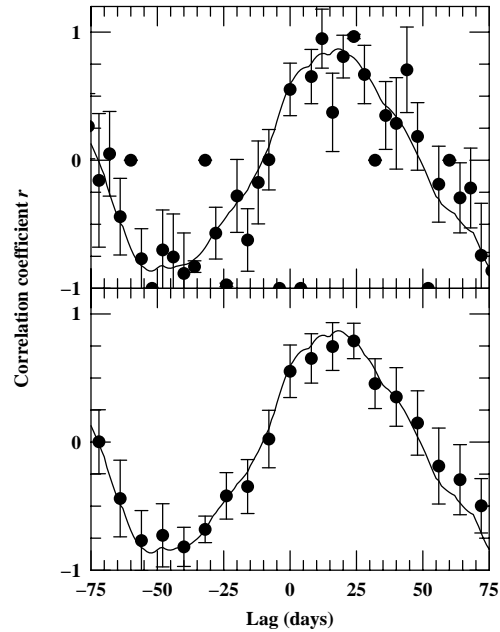,width=7.5cm,angle=0}}
  \end{center}
\caption{Cross-correlation functions for the
Mrk 335 light curve shown in Fig.~\ref{fig:timecor}.
The DCF values are shown as points with error bars,
and the interpolation CCF, as in Fig.~\ref{fig:ccf},
is shown as a solid line. The upper panel shows the
DCF with a bin width of $\Delta \tau = 4$ days,
and the lower panel shows a bin width of 
$\Delta \tau = 8$ days; with $\Delta \tau = 4$ days,
some bins have no data (those with values set to
$r=-1$), but with $\Delta \tau = 8$ days, the DCF is
somewhat underresolved.
\label{fig:dcf}}
\end{figure}

There are some circumstances under which one might not
be able to reasonably interpolate between gaps in data.
This can occur (a) when there are a few large gaps in
otherwise well-sampled series or (b) when there is reason
to believe that the variations might be at least somewhat
undersampled. In these cases, interpolation might be
highly misleading, and another methodology needs to be
employed. The ``discrete correlation function'' (DCF)\cite{EdKr88}
method is one where no assumption about light curve behavior
needs to be made. The DCF method deals with irregularly
sampled data by binning the data in time, as illustrated
schematically in Fig.~\ref{fig:dcfbins}. This is an
alternative approach to the irregular sampling requirement:
instead of requiring that points contributing to 
${\rm CCF}(\tau)$ are separated in time by exactly
the interval $\tau$, we time-bin the data by pairing
points with time separations in the range
$\tau \pm \Delta \tau/2$, where $\Delta \tau$ is
the width of one time bin. Choice of the
binning window is a free parameter, and two examples
are shown in Fig.~\ref{fig:dcf}.

The principal virtues of the DCF method are (a) 
that only actual data points are used and (b) that it
is possible to assign a statistical uncertainty to the value of
the correlation coefficient in each bin. The relative
weakness of the DCF is that the data are in some ways underutilized,
as is evident in Fig.~\ref{fig:dcfbins}; for a small data set,
the DCF method might completely miss a real correlation,
although it is less like to find a spurious correlation
than is the interpolation method\cite{WhPe94}.

One difficulty of the DCF method is that
the number of points per time bin can vary
greatly, as can be easily inferred from inspection of 
Figs.~\ref{fig:dcfbins} and \ref{fig:dcf}.
One solution to this is to vary the width of
the time bins to ensure that there are a statistically
meaningful number of points in each bin. A method for
accomplishing this is the ``Z-transformed DCF (ZDCF)''\cite{Al97}.

\subsection{Computational Issues}

In the foregoing discussion, we have concentrated primarily
on conceptual issues without paying much attention to a 
number of computational issues that must be dealt with
in actually computing CCFs\cite{WhPe94}$^,$\cite{We00}$^,$\cite{ViWa00}.

\begin{enumerate}
\item {\em Edge effects.} Only at zero lag do all the points
of the time series enter into the calculation of the
correlation coefficient; at any other lag, points at
the ends of the time series drop out of the calculation
since there are no points in the other series to 
which they can be matched (see Fig.~\ref{fig:interplc}).
This means that at larger and larger lags, fewer and
fewer points contribute to the calculation of $r$.
This has two significant consequences:
\begin{enumerate}
\item {\em Normalization.} Correct normalization of the 
CCF requires use of the correct value of the mean and
standard deviation for each series, as can be seen from
the definition of the correlation coefficient $r$
(Eq.~(\ref{eq:rdef})). Since AGN time series are limited
in duration, as points near the edges drop out of the
calculation, the mean and standard deviation of the
series change; in statistical language, this means
that the series are ``non-stationary.'' Thus, the mean
and standard deviation of the series needs to be
recalculated for every lag, using only the data points
that actually contribute to the calculation. In the
case where interpolated values from one series are
used, the mean and standard deviation should be those
of the interpolated points, not the original points.
\item {\em Significance.} Once the peak value of the
CCF $r_{\rm peak}$ has been found, we want to know
whether or not the correlation found is 
``statistically significant'', i.e., is it likely
to be real or spurious? The statistical significance
of the correlation depends on the number of points
that contribute to the calculation of $r_{\rm peak}$,
not to the total number of points in the series.
Suppose, for example, that we have two time series
consisting of $N=50$ points, but that the maximum
value of the CCF (say, $r_{\rm peak} = 0.45$)
occurs at a such a large lag that only $N=30$ of the points are actually
contributing at this lag. For a linear correlation 
coefficient of $r = 0.45$ and $N=30$, the correlation is
significant at about the 98\% confidence level. However,
if we erroneously use $N=50$, we would conclude that the level of
significance is about 99.9\%, clearly a major difference.
If we fail to account for the correct number of
points contributing to the calculation of $r_{\rm peak}$
and simply use the number of points in the series, we will
{\em overestimate} the significance of the correlations we
detect.
\end{enumerate}
\item {\em Interpolation scheme.} There are a couple of
issues that arise in this regard:
\begin{enumerate}
\item {\em Which series?} In the examples shown in
Fig.~\ref{fig:interplc}, we have interpolated in the
emission-line light curve. Is there any particular reason
to choose one series over the other when doing the interpolation?
In general there
does not seem to be, unless, for example, the emission-line
light curve is markedly smoother than the continuum light
curve (on account of the time-smearing effect) or one light
curve is much better sampled than the other. Usually,
one computes the CCF by computing the CCF twice, interpolating
once in each series, and then averaging the results\cite{GaPe87}.
\item {\em Interpolating function.} Here we have considered
only point-to-point linear interpolation, which results in
first-derivative discontinuities in the light curves and
the CCFs. Is there any advantage to using higher-order 
interpolation functions in the time series? Generally, no,
higher-order functions don't improve the CCFs in any sense,
and can be grossly misleading, as higher-order fits based
on only a few data points become hard to control.
\end{enumerate}
\item {\em Resolution of the CCF.} Can the accuracy of
a cross-correlation result be better than the typical
sampling interval? Yes, it can, as long as the functions
involved are reasonably smooth. The analogy that is usually
drawn is that one can measure image centroids to far higher
accuracy than the image size, which is true because
both stellar images and instrumental point-spread functions
are generally smooth and symmetric. Statistical tests as
described below suggest that uncertainties of about 
half the sampling interval are routinely obtained.
\end{enumerate}

\subsection{Uncertainties in Cross-Correlation Lags}

Although cross-correlation techniques have been applied to
AGN time series for about 15 years, there is still no obvious
or even universally agreed-upon way to assess the uncertainties
in the lag measurements obtained. At present, the most effective
technique seems to be a model-independent Monte-Carlo method
known as FR/RSS (for ``flux redistribution/random subset selection'')\cite{Pe_ea98b}.

FR/RSS is based
on a computationally intensive statistical method known as
a ``bootstrap''. The bootstrap works as follows: suppose that
you have a set of $N$ data pairs ($x_i, y_i$) and that linear
regression yields a correlation coefficient $r$. How accurate
is $r$? In particular, how sensitive is it to the influence of
individual points? One can assess this by a Monte Carlo process
where one selects at random $N$ points from the original
sample, without regard to whether or not any point has been
selected previously. For the new sample of $N$ points (some of
which are redundant selections from the original sample, while
some points in the original sample are missing), the linear
correlation coefficient is recalculated. When this is done many
times, a distribution in $r$ is constructed, and from this, one
can assign a meaningful statistical uncertainty to the original
experimental value of $r$.

This process can also be assigned to time series, except that
the time tags of the points have to be preserved. In effect, then,
this means that redundant selections are overlooked; the
probability that in $N$ selections of $N$ points a point will
be selected zero times is $1/e$, so the new time series, selected
at random, has typically fewer points by a factor of $1/e$
(hence the name ``random subset selection''). Welsh\cite{We00} suggests
that this should be modified in the sense that the weighting
of each selected point should be proportional to $\sqrt{n_i}$,
where $n_i$ is the number of times the data point
$(x_i, y_i)$  is selected in a single realization.
This is closer in philosophy to the original bootstrap, but
it has not been rigorously tested yet.

The other part of the process, ``flux redistribution,''
consists of changing the actual observed fluxes in a way
that is consistent with the measured uncertainties.
Each flux is modified by a random Gaussian deviate based
on the quoted error for that datum (i.e., after a large
number of similar modifications, the distribution of
flux values would be a Gaussian with mean equal to the
data value and standard deviation equal to the quoted error).
\begin{figure}
\begin{center}
    \leavevmode
  \centerline{\epsfig{file=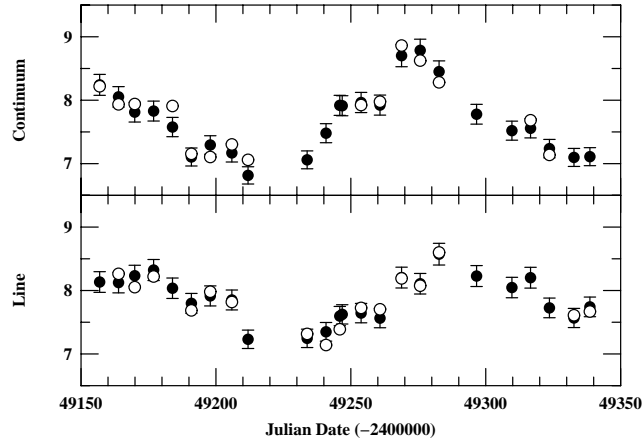,width=8.5cm,angle=0}}
  \end{center}
\caption{The filled circles show the Mrk 335 light curve
from Fig.~\ref{fig:timecor}, but now
with error bars shown. The open circles show a
single FR/RSS Monte Carlo realization; the points are
selected from the total subset at random, and the
fluxes are adjusted as Gaussian deviates. Note that some
of the original data points are not seen because the
selected flux-adjusted points cover them.
The realization shown here gives $\tau_{\rm cent}
=17.9$ days, compared to $\tau_{\rm cent} = 15.6$ days for
the whole data set.
\label{fig:frrss}}
\end{figure}
\begin{figure}
\begin{center}
    \leavevmode
  \centerline{\epsfig{file=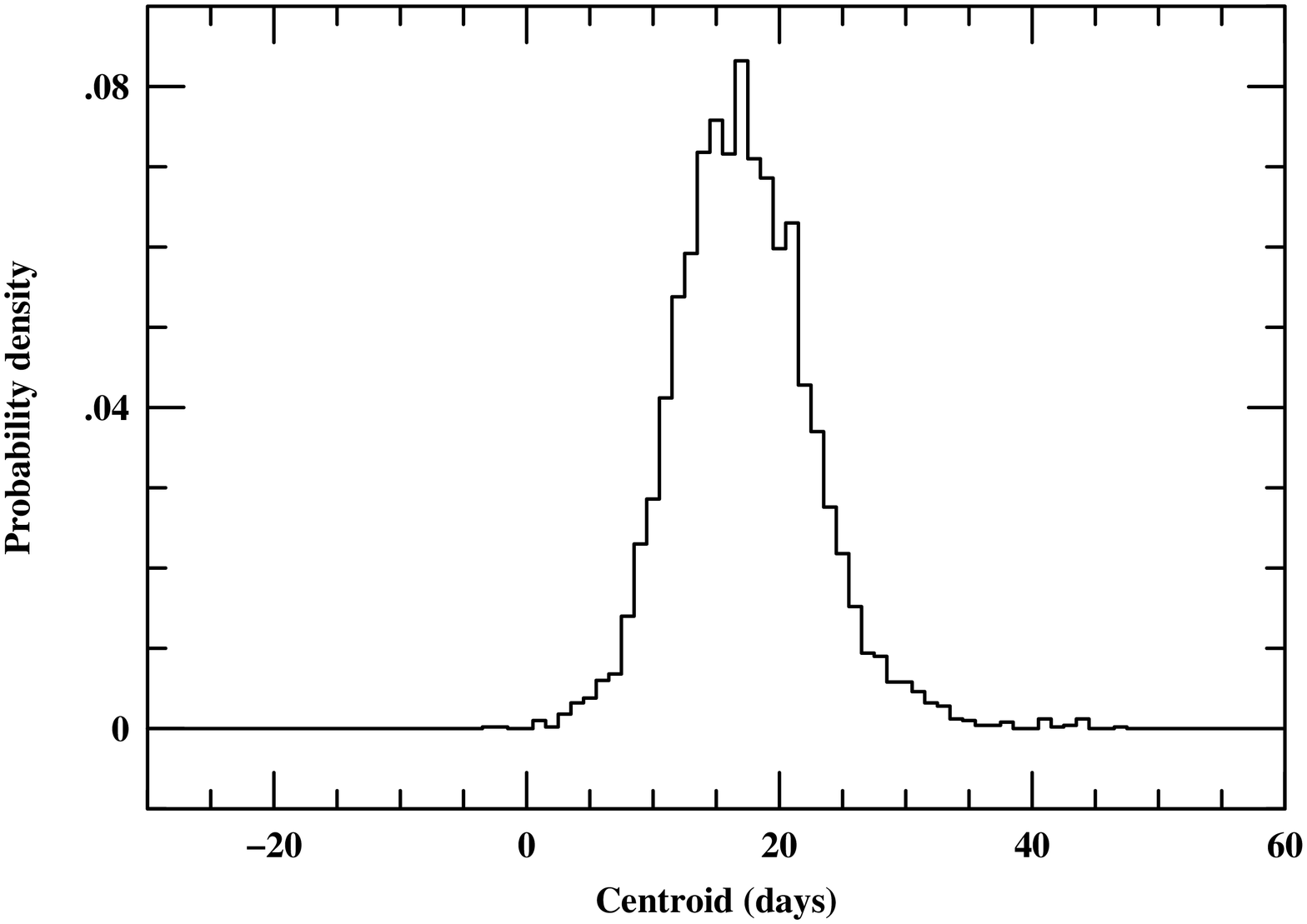,width=8cm,angle=0}}
  \end{center}
\caption{Multiple Monte-Carlo realizations such as
those in Fig.~\ref{fig:frrss} are used to build up
a cross-correlation peak distribution (CCPD). 
Relative to the measured CCF centroid
for the whole data set ($\tau=15.6$ days), the $\pm1\sigma$ 
width of this distribution is $+7.2$, $-3.1$ days.
\label{fig:ccpd}}
\end{figure}

A single sample FR/RSS realization is shown schematically
in Fig.~\ref{fig:frrss}. For each such realization,
a cross-correlation is performed and the 
centroid is measured. A large number of similar realizations
will produce a ``cross-correlation peak distribution''
(CCPD)\cite{MaNe89}, as shown in Fig.~\ref{fig:ccpd}. The CCPD
can be integrated to assign formal 
uncertainties (usually $\pm1\sigma$) to the value of $\tau_{\rm cent}$
measured from the entire data set.

\section{Observational Results}
\label{sect:results}

With the background provided in the previous sections, we can
discuss some of the more important observational results.
We will begin with emission-line variability, since the
results obtained thus far have been relatively less ambiguous
than the results on interband continuum variations.

\subsection{Size of the Broad-Line Region}

\begin{figure}
\begin{center}
    \leavevmode
  \centerline{\epsfig{file=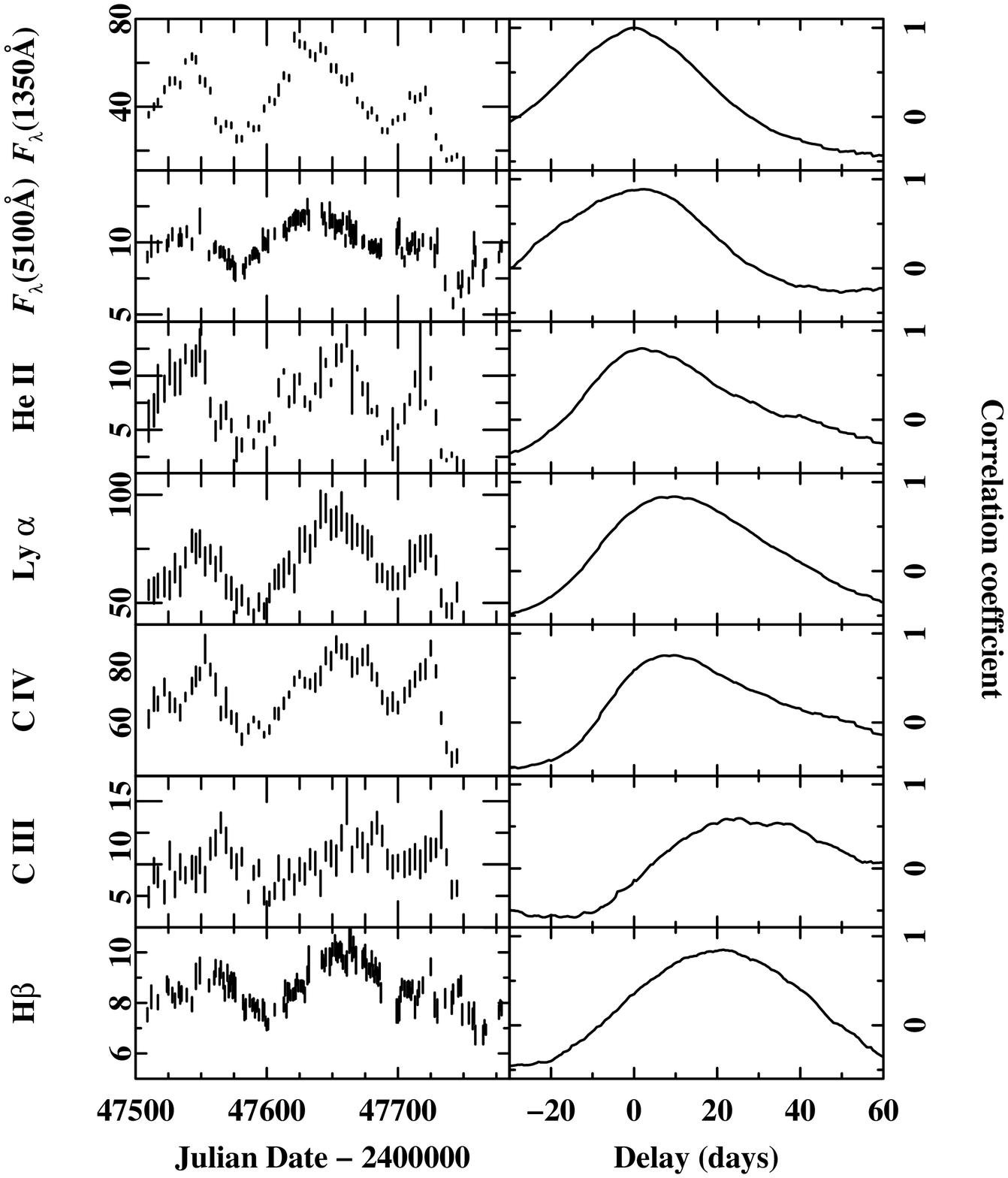,width=11cm,angle=0}}
  \end{center}
\caption{Light curves (left-hand column) and
CCFs (right-hand column) for NGC 5548, from the first International
AGN Watch monitoring project in 1988--89.
From the top, the light curves are
the UV continuum at 1350\,\AA,
the optical continuum at 5100\,\AA,
\heii\,$\lambda1640$,
\Lya\,$\lambda1215$,
\civ\,$\lambda1549$,
\ciii]\,$\lambda1909$, and
\Hbeta\,$\lambda4861$. The continuum fluxes are in units
of $10^{-15}$\,\efluxA, and the line fluxes are in units
of $10^{-13}$\,\ergscm2. The CCFs are computed relative to the 
UV continuum at the top; the top panel is the UV continuum ACF.
Data from Clavel et al.$^{14}$ and Peterson et al.$^{67}$.
\label{fig:ngc5548}}
\end{figure}

By the late 1980s, the potential power of reverberation
mapping was widely understood, but the high-intensity
monitoring programs required to extract this information
had not been carried out, mainly on account of sociological
barriers: the sampling requirements, in terms of time
criticality and number of observations, were far in
excess of anything that had been done in the past on
oversubscribed shared facilities. In late 1987, a large
informal consortium known as the {\em International
AGN Watch}\footnote{More information about the International
AGN Watch and all papers and data obtained are available on
the AGN Watch website at
{\sf http://www.astronomy.ohio-state.edu/$\sim$agnwatch/.}} was 
formed in an attempt to obtain sufficient
observing time with IUE and various ground-based
observatories to measure emission-line lags for the
Seyfert 1 galaxy NGC 5548, even at that time one of the best-studied
variable AGNs. This turned out to be an enormously
successful project\cite{Cl_ea91}$^,$\cite{Pe_ea91}$^,$\cite{Pe_ea92}
that provided impetus for
additional similar projects by the AGN Watch
and other groups (see Peterson\cite{Pe93} for a review
of the field in the wake of the first monitoring
project on NGC 5548). Some of the light curves
and CCFs from the original NGC 5548 project are shown
in Fig.~\ref{fig:ngc5548}.

Since this initial successful campaign, a handful of
AGNs have been well-monitored in the UV and optical
(NGC 3783, NGC 4151, NGC 7469, 3C 390.3, Fairall 9, in
addition to NGC 5548) and nearly three dozen Seyfert
galaxies and low-luminosity quasars have been closely
monitored in the optical\cite{WPM99}$^,$\cite{Kas_ea00}.
The most important general conclusions reached from these monitoring
programs are:
\begin{enumerate}
\item The UV/optical continua vary together, to within a
couple of days.
\item The highest ionization emission lines respond most rapidly
to continuum changes,
implying that there is  ionization stratification of the
BLR.
\item The BLR gas is not in pure radial motion, since no 
unambiguous
differences in the time scale of the response of the blue
and red wings of the emission lines have been detected.
\end{enumerate}

\begin{figure}
\begin{center}
    \leavevmode
  \centerline{\epsfig{file=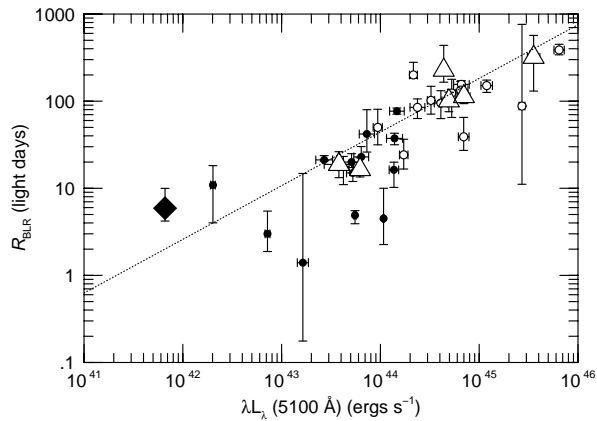,width=8.5cm,angle=0}}
  \end{center}
\caption{The BLR radius, as measured by the \Hbeta\ lag,
as a function of average optical continuum luminosity.
The open symbols are Palomar--Green quasars from
Kaspi et al.$^{36}$, the filled circles are Seyfert 
1 galaxies from Wandel et al.$^{88}$.
The large symbols are narrow-line 
($\vFWHM < 2000$\,\kms) AGNs, and the filled
diamond is NGC 4051$^{75}$.
The dotted line is the best fit to all the data 
except NGC 4051.
\label{fig:radlum}}
\end{figure}

Cross-correlation results
show that the size of the \Hbeta-emitting region scales
with average continuum luminosity as\cite{Kas_ea00}
\begin{equation}
\label{eq:AGNrl}
r(\Hbeta) \approx 0.22
\left( \frac{L_{\rm bol} \ho^2}{10^{46}\,\ergsec} \right)^{0.7}\ {\rm pc},
\end{equation}
 and that, in general, sizes derived from other lines
usually scale as 
$r (\mbox{\rm \heii, \nv}) \approx 0.2 r (\Hbeta)$ and
$r (\mbox{\rm \Lya, \civ}) \approx 0.5 r (\Hbeta)$. 
A slightly different version of the 
BLR radius-luminosity relationship is shown in
Fig.~\ref{fig:radlum}, in which $r \propto L^{0.6\pm0.1}$.
This is close to the slope of the relationship predicted by
the most naive sort of scaling: since to lowest order AGN spectra
all look alike, the ionization parameters (Eq.~(\ref{eq:Udef}))
and particle densities must be the same in all of them.
Thus, inverting Eq.~(\ref{eq:Udef}), we predict that
\begin{equation}
\label{eq:radlum}
r  =  \left( \frac{Q({\rm H})}{4 \pi c n_{\rm e}} \right)^{1/2}
 \propto  L^{1/2} \mbox{\ \ ,}
\end{equation}
where in the last step we have also assumed that
the shape of the ionizing continuum is not a function of
luminosity. Despite these gross assumptions, 
the agreement with the data is not bad, although the
large scatter alone tells us that this is not the
whole story.

\subsection{AGN Black-Hole Masses}

\begin{figure}
\begin{center}
    \leavevmode
  \centerline{\epsfig{file=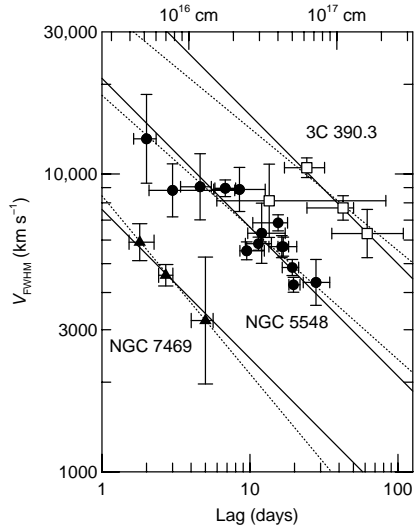,width=5.5cm,angle=0}}
  \end{center}
\caption{Line width in the
rms spectrum plotted as a function of the distance from the central source
(upper horizontal axis) as measured by the emission-line lag 
(lower horizontal axis) for various broad emission lines in 
NGC~7469, NGC~5548, and 3C 390.3. The dashed lines are 
best fits of each set of data to the relationship
$\log \vFWHM = a + b\log c\tau$.
The solid line shows the best fit to each set of data for fixed $b=-1/2$,
yielding virial masses of 
$8.4\times 10^6\,\Msun$, 
$5.9\times 10^7\,\Msun$, and 
$3.2\times 10^8\,\Msun$ for the three respective galaxies.
From Peterson \& Wandel$^{72}$ \copyright 2000 AAS..
\label{fig:virial}}
\end{figure}

Reverberation mapping is one method of measuring AGN central 
masses via the virial relationship
\begin{equation}
\label{eq:virial}
M = \frac{f r \sigma^2}{G} \mbox{\ \ ,}
\end{equation}
where $f$ is a dimensionless factor of order unity that
depends on the geometry and kinematics of the BLR,
$\sigma$ is the emission-line velocity dispersion,
and $r$ is the size of the emitting region. 
Measurement of the emission-line time lags
provides the ingredient that has been missing since
the first attempts to understand the basic physics of
AGNs\cite{Wo59}. Relative to other dynamical
estimators, advantage of using the BLR to
provide an estimate of the mass of the central source
is that it is located very close to the central
source (within $\sim10^3 R_{\rm grav}$), leaving little
doubt that the central mass is in fact a black hole. On
the other hand, the kinematics of the BLR are not yet
understood (see below), and non-gravitational forces
might have a strong effect on gas motions.
For the virial method to be applicable, 
the BLR kinematics must be dominated by gravity.
Even without understanding the detailed geometry
and kinematics of the BLR, we can test the virial
hypothesis by  comparing lags and line widths 
measured in a single AGN:
all lines must give the same virial mass, even though
not all the line-emitting material needs to have 
common kinematics. The three AGNs for which this
can be easily tested are shown in Fig.~\ref{fig:virial}.
\begin{figure}
\begin{center}
    \leavevmode
  \centerline{\epsfig{file=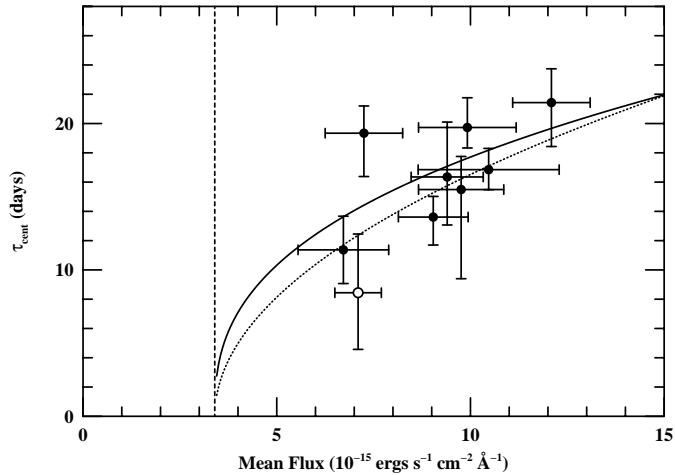,width=10cm,angle=0}}
  \end{center}
\caption{The \Hbeta\ emission-line lag 
as a function of optical continuum flux for the Seyfert 1
galaxy NGC 5548. The solid line shows the best power-law
fit to the data, $\tau \propto F^{0.4\pm0.2}$, and
the dotted line shows the best fit to the naive theoretical
prediction $\tau \propto F^{1/2}$ (Eq.~(\ref{eq:radlum})).
The dashed vertical line shows the estimated starlight
contribution, as in Fig.~\ref{fig:ngc5548lc}.
\label{fig:lumtau}}
\end{figure}
\begin{figure}
\begin{center}
    \leavevmode
  \centerline{\epsfig{file=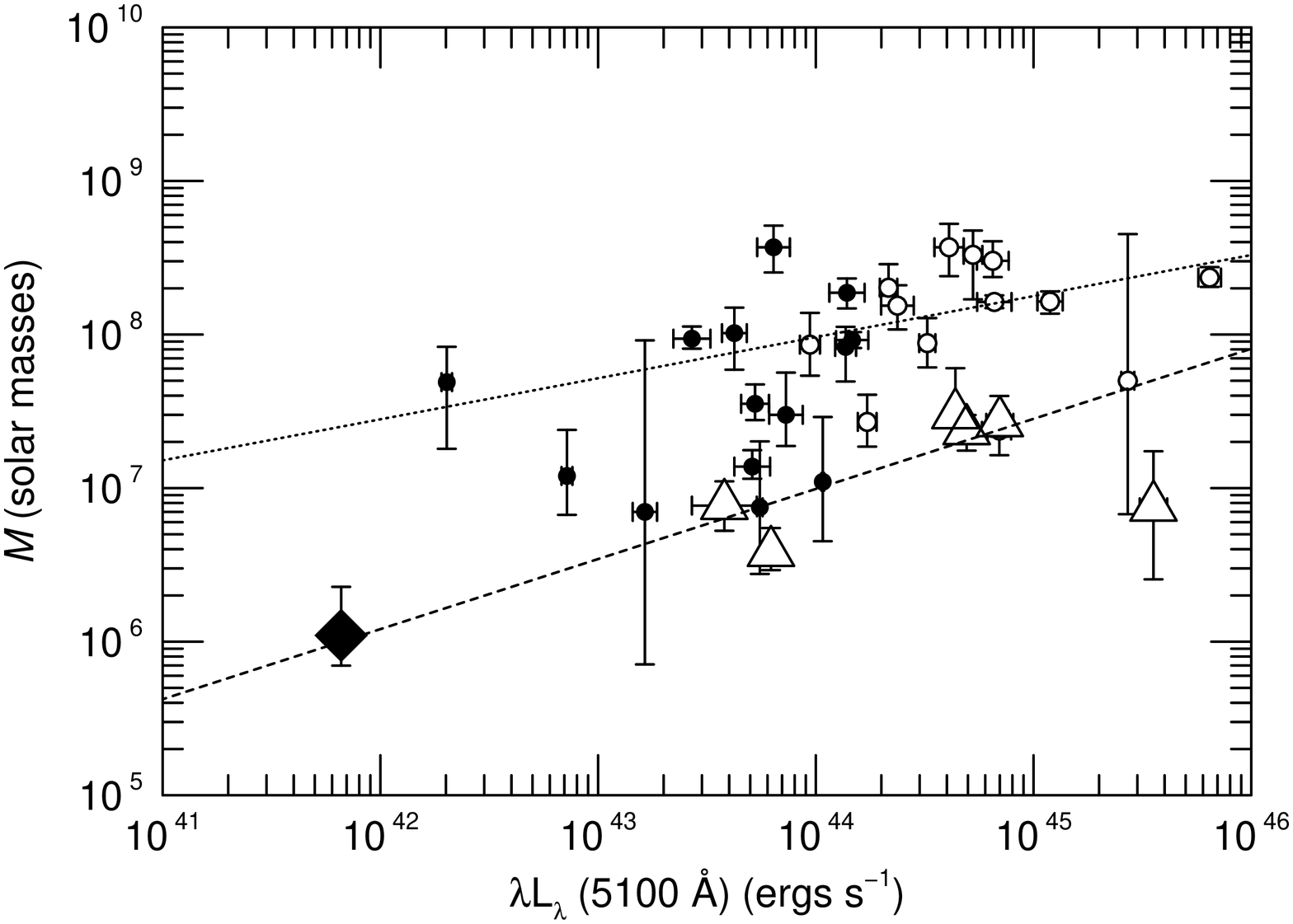,width=9cm,angle=0}}
  \end{center}
\caption{The central mass of AGNs, as measured 
from \Hbeta\ 
line 
reverberation,
as a function of optical continuum luminosity.
Symbols are as in Fig.~\ref{fig:radlum}.
From Peterson et al.$^{75}$ \copyright 2000 AAS.
\label{fig:masslum}}
\end{figure}

Even if this were not true for all lines, it may be 
true for some lines, and
a given line must always yield the same mass. Only
in the case of NGC~5548 is there sufficient information
on the long-term behavior of a single line (\Hbeta)
for this test to be applied, and the data seem to
be consistent with the virial relationship\cite{PeWa99}.
We expect, then, that
as the continuum brightens, the emission-line
lag increases (see Fig.~\ref{fig:lumtau}), 
and the emission-line becomes narrower.
This does seem to be the case.

Virial masses based on \Hbeta\ line reverberation 
as a function of optical luminosity are shown
in Fig.~\ref{fig:masslum}. There is considerable
scatter in the relationship, but it is nevertheless
clear that higher-mass black holes are found in
higher-luminosity AGNs. Some of the scatter in
this relationship may be attributable to differences
in accretion rate or radiative efficiency: the
lower end of the envelope, for example, seems to
be dominated by narrow-line Seyfert 1 galaxies,
which are thought to have relatively high accretion
rates (and thus luminosities) for their mass.
Additional factors, such as inclination of the system,
may also contribute to the scatter. But we are,
finally, beginning to see the first indications 
of a mass-luminosity relationship for AGNs.

\subsection{Emission-Line Transfer Functions}

While existing AGN monitoring data have been of sufficient
quality and quantity to obtain cross-correlation
lags, recovery of transfer functions has, not surprisingly,
proven to be far more difficult. Existing
transfer function solutions tend to be very noisy
and ambiguous.

\begin{figure}
\begin{center}
    \leavevmode
  \centerline{\epsfig{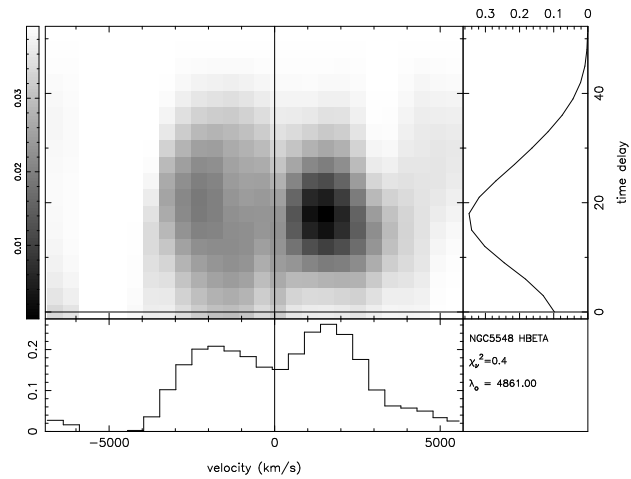}}
  \end{center}
\caption{A transfer function for \Hbeta\ in NGC 5548,
based on AGN Watch data from 1988--1996.
Courtesy of S.\ Collier.
\label{fig:hbmap}}
\end{figure}

In Fig.~\ref{fig:hbmap}, we show a sample transfer
function for the \Hbeta\ emission line in NGC 5548;
this transfer function is based on data extending
over more than a crossing time, so the reader is
cautioned against concluding too much from this example,
since it is based on data that span a long  interval 
compared to the BLR dynamical time scale
(Eq.~(\ref{eq:dynamical})).
The structures seen in this transfer function
do not correspond to those seen
in any of the simple models that we have described
earlier. Note in the one-dimensional transfer function
the low amplitude of the response
at zero lag, first noticed even with just the
first year of AGN Watch data\cite{HWP91};
this indicates that there is little response due to material
along our line-of-sight to the continuum source, suggestive
of either a low-inclination disk (i.e., there is little
gas along the line of sight) or anisotropic line response\cite{Fe_ea92}.
Whether or not other lines have
small response at small lag is less certain\cite{Kr_ea91}$^,$\cite{Pe93}; 
this effect
may be seen clearly only in \Hbeta\ on account of the
large lag.

\begin{figure}[t]
\begin{center}
    \leavevmode
  \centerline{\epsfig{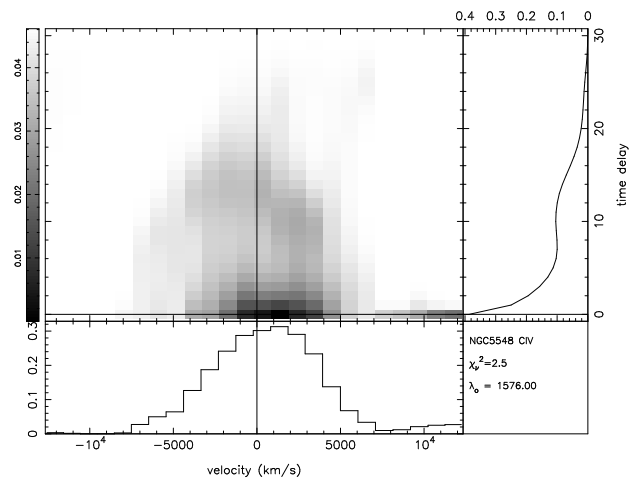}}
  \end{center}
\caption{A transfer function for \civ\,$\lambda1549$ in NGC 5548,
HST observations$^{40}$ from 1993.
Courtesy of S.\ Collier.
\label{fig:c4map}}
\end{figure}

Fig.~\ref{fig:c4map} shows an attempt to recover the
\civ\ transfer function from 39 daily observations 
of NGC 5548 made with HST in 1993. These data have
been used in several independent analyses\cite{Wa_ea95}$^,$\cite{DoKr96}$^,$\cite{ChMu96}$^,$\cite{Bo_ea97}
with no consensus on the
interpretation, and indeed with quite contrary conclusions
about the kinematics: Wanders et al.\cite{Wa_ea95} favor no radial
motion, Done \& Krolik\cite{DoKr96} find a hint of radial infall
(also previously suggested by Crenshaw \& Blackwell\cite{CrBl90}
on the basis of the first year of IUE data),
and Chiang \& Murray\cite{ChMu96} and Bottorff et al.\cite{Bo_ea97}\ fit 
the data with 
different radial-outflow models.

A crude measure of the velocity field might be obtained 
by cross-correlating parts of the emission line:
for example, comparison of the response times for the
red and blue wings could be used to detect
radial  infall/outflow. Similarly, one could
compare the response times for the line wings with
the line core in an attempt to detect virial
motion, i.e., $V \propto r^{-1/2}$.
There have been a number of reports that the 
emission lines wings respond faster than 
core\cite{Cl91}, as expected for virial motion, but detection is 
always weak.

\begin{figure}
\begin{center}
    \leavevmode
  \centerline{\epsfig{file=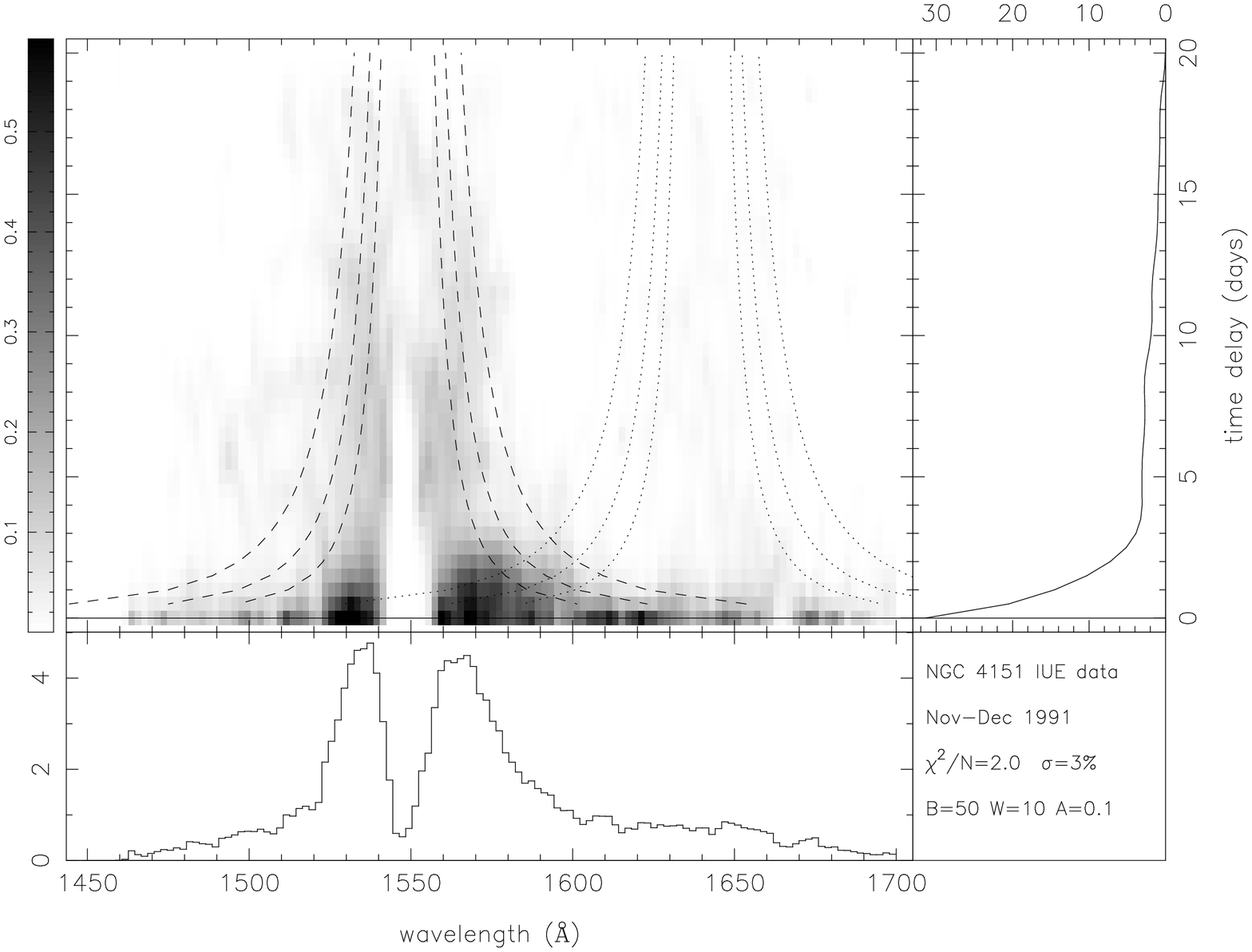,width=6.2cm,angle=-90}}
  \end{center}
\caption{A transfer function for the \civ--\heii\
region of NGC 4151 based on IUE spectra. The strong double-peaked
appearance of the \civ\ line is due to a deep 
slightly blueshifted absorption feature in this line.
From Ulrich \& Horne$^{85}$ \copyright 1996 Blackwell Sci.
\label{fig:4151tf}}
\end{figure}

A two-dimensional transfer function for the
\civ--\heii\ spectral region in  NGC 4151
is shown Fig.~\ref{fig:4151tf}.
Ulrich \& Horne\cite{UlHo96} argue that there is a hint
that the red wing responds slightly more rapidly
than the blue wing, which is the expected signature
for radial infall.


\subsection{Emission-Line Profile Variability}

\begin{figure}
\begin{center}
    \leavevmode
  \centerline{\epsfig{file=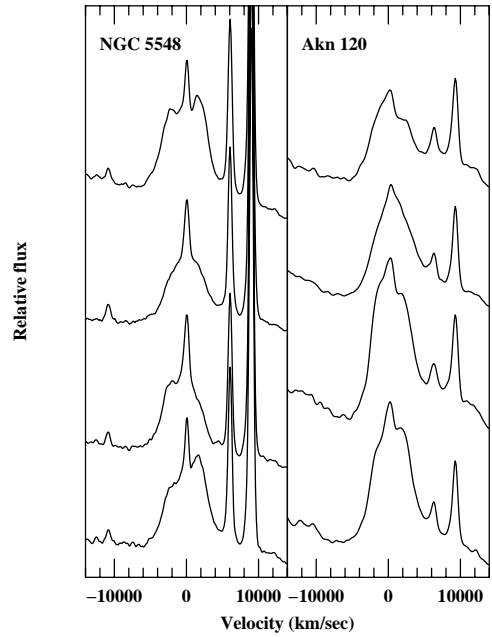,width=6.5cm,angle=0}}
  \end{center}
\caption{Emission-line profiles for \Hbeta\ in
NGC 5548 (left-hand column; International AGN Watch data) and 
Akn 120 (right-hand column; data from Peterson et al.$^{69}$).
These are representative profiles obtained over several
years of monitoring. The \Hbeta\ narrow-line component is particularly
noticeable in the spectra of NGC 5548.
\label{fig:profiles}}
\end{figure}

If the BLR velocity field is ordered (i.e., $V = f (r)$), then
the emission-line profiles should vary in response to continuum changes:
the propagation of excitation inhomogeneities through the
BLR should result in irregularities in line profiles that evolve
on time scales comparable to the emission-line lags.
Peterson et al.\cite{PPW99} have examined profile
variations in several Seyfert galaxies that were 
monitored fairly intensively over several years. They
conclude that profile variations are not obviously correlated with 
either the continuum or the total emission-line flux;
profile variations are  not reverberation effects.
They also find that most \Hbeta\ profiles can be modelled
fairly well with three Gaussian components that are
fixed in location and width and vary only in relative
flux, which thus allows multiple-peaked profiles
and variable asymmetry. 
They also conclude that profile irregularities can occur quickly,
appearing on time scales as short as days or weeks, but they
can then persist for a very long time, months or even years.
Sample \Hbeta\ profiles for two well-studied Seyferts are shown
in Fig.~\ref{fig:profiles}.

\subsection{Multiwavelength Continuum Variability}

As noted earlier, one of the important conclusions
reached from the original AGN Watch monitoring
program was that the UV and optical continua
vary with no apparent time delay between them,
at least to the accuracy of the experiment
($\ltsim 1$ day in the best cases). An important
consequence of this is that AGN continuum variability
is not due either to mechanical accretion-disk instabilities
or to variations in the accretion rate,
since such effects would be expected to propagate through
the disk (and thus across the spectrum) on 
much-longer sound-crossing (Eq.~(\ref{eq:sound})) or
drift time scales (Eq.~(\ref{eq:drift})), respectively.
The suggestion was made\cite{Co91}$^,$\cite{Cl_ea92}
that the UV/optical variations might in fact be
driven by X-ray variations, possibly in a manner
consistent with the X-ray reprocessing models
that were being developed to account for the
10 keV reflection hump and the equivalent width
of the 6.4\,keV Fe K$\alpha$ 
line\cite{GuRe88}$^,$\cite{LiWh88}$^,$\cite{GeFa91}; both of these
features suggest that something like half of the
emitted hard X-rays in AGN interact with
``cold'' (not highly ionized) matter that covers much
of the sky as seen from the source (and which might
in fact be the accretion disk itself).
If it is supposed that hard X-rays are produced
above the disk plane near the axis of the accretion
disk, then hard X-ray radiation striking the disk
should be reprocessed into
lower-energy continuum photons. Moreover, 
the UV/optical variations should follow those in
the X-rays, with the shortest time delays for
the higher-energy photons that are produced
predominantly in the central regions of the
accretion disk. The temperature structure of
a classical thin accretion disk is given by
\begin{equation}
T^4 = \frac{3 G M \dot{M}}{8 \pi \sigma r^3}
\left[ 1 - \left( \frac{r}{R_{\rm min}} \right)^{-1/2} \right] \mbox{\ \ ,}
\end{equation}
and assuming that AGNs are accreting close to the Eddington rate
with about 10\% efficiency, the radial temperature
structure becomes
\begin{equation}
T(r)  \approx 6.3 \times 10^5
\left( \frac{\dot{M}}{\dot{M}_{\rm Edd}} \right)^{1/4}
\left( \frac{M}{10^8\,\Msun} \right)^{-1/4}
\left( \frac{r}{R_{\rm S}} \right)^{-3/4}\ {\rm K} \mbox{\ \ ,}
\end{equation}
from which we expect a peak in the spectral energy distribution 
in the UV/soft X-ray region, which as noted earlier might
in fact be the origin of the big blue bump\cite{Sh78}$^,$\cite{MaSa82}.

If we suppose that a thin accretion disk is irradiated
by an X-ray source on the disk axis, we should see the
inner, hotter part of the accretion disk respond before
the outer, cooler parts. 
From Wien's Law, the
radiation at wavelength $\lambda$ arises primarily at
a particular temperature corresponding to a particular
location in the disk, i.e.,
$\lambda \propto T^{-1} \propto r^{3/4}$, so the
difference in the response times at different
wavelengths should be naively
\begin{equation}
\Delta \tau \propto \lambda^{4/3} \mbox{\ \ .}
\end{equation}
It is straightforward to show that this relationship
holds for an irradiated disk that is heated locally
by hard X-rays. 

\begin{figure}
\begin{center}
    \leavevmode
  \centerline{\epsfig{file=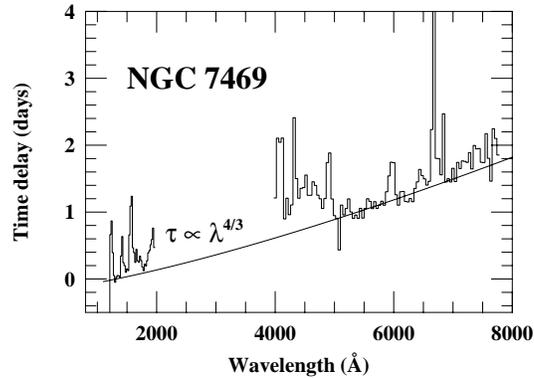,width=7cm,angle=0}}
  \end{center}
\caption{Lag as a function of  wavelength for NGC 7469. The average flux in
each wavelength band is cross-correlated with the 
shortest-wavelength UV (1315\,\AA)
continuum, yielding a ``lag spectrum''. The expected dependence 
for an irradiated 
thin accretion disk, $\tau \propto \lambda^{4/3}$, is also shown.
The broad emission lines are prominent because they have longer
response times than the adjacent continuum.
\label{fig:7469lag}}
\end{figure}

The predicted wavelength-dependent continuum time
delays have been detected reliably in the case
of NGC 7469, which was monitored intensively
in 1996\cite{Wa_ea97}$^,$\cite{Co_ea98}$^,$\cite{Kr_ea00}.
Fig.~\ref{fig:7469lag} shows
the result of measuring the UV and optical fluxes across
the entire monitored UV and optical spectrum, and
cross-correlating each of the resulting light curves
with the shortest-wavelength UV continuum. 
Relative to the short-wavelength UV, continuum lags
in other line-free continuum bands are
detected at no less than 97\% confidence 
throughout the UV/optical region.

While the evidence shows that there are in fact wavelength-dependent
continuum lags in NGC 7469, this is not an unambiguous detection
of a classical thin accretion-disk structure. There are several complications:
\begin{enumerate}
\item Similar wavelength-dependent lags have not been reliably 
detected in other sources. This effect may be marginally
present in the case of NGC 4151\cite{Pe_ea98b},
but it is not found in NGC 3516\cite{Ed_ea00}.
However, the sampling rates obtained for
NGC 7469 are far superior to those of most other
experiments, and the upper limits on UV/optical continuum
lags for other AGNs seem to be consistent with predictions
based on (highly uncertain) scaling of the other sources
to the NGC 7469 result, except possibly in the case of
NGC 3516.
\item Korista \& Goad\cite{KoGo00} have suggested 
wavelength-dependent lags are consistent with the 
expectations for diffuse emission from the emission-line 
clouds (Balmer and Paschen continuum). Observations over
a wider wavelength range are needed to
distinguish between an accretion-disk and
a diffuse-continuum origin (specifically, the NGC 7469 observations
did {\em not} cover the crucial wavelength range around
the Balmer edge $\sim3650$\,\AA).
\item The hard X-ray flux variations in NGC 7469 are not 
correlated with those in the UV/optical\cite{Na_ea98}.
The hard X-ray spectrum was monitored with RXTE
at the same time as the UV/optical monitoring campaign.
Comparison of the light curves shows that the minima
occurred at about the same time, but the hard X-ray
maxima occurred approximately 4 days {\em later}
than the corresponding maxima in the UV continuum light curve.
This is a {\em prima facie} argument against the reprocessing 
scenario. However, a more complete spectral analysis\cite{Na_ea00} 
enabled by
an improved background model for RXTE 
finds that the UV flux variations are correlated with 
variations in the X-ray spectral index, in the sense
that the X-ray spectrum is {\em softer} when the
UV is brighter. This suggests that the soft X-rays are 
strongly correlated with the UV; an increase in
the number of UV seed photons cools the Comptonizing corona
above the accretion disk, softening the spectrum.
\end{enumerate}

By way of contrast,
a recent study of high-energy variability
in NGC 5548 shows clear relationships between the hard X-rays
and the lower-energy photons\cite{Ch_ea00}:
the extreme ultraviolet (EUV) variations lead those
in the soft X-ray region by about 3.5 hrs and those
in the hard X-ray by about 10 hrs. The temporal order
and time scales are consistent with production of
both the soft and hard X-rays by 
Compton upscattering of lower-energy 
photons.

If the X-rays are indeed Comptonized lower-energy photons,
are the seed photons in the EUV ($\sim100$\,\AA)
or UV ($\sim1000$\,\AA)?
The only good comparison available is based on
simultaneous UV and EUV 
monitoring of NGC 5548 in 1993\cite{Ma_ea97}.
Cross-correlation of the
UV and EUV light curves shows that the UV leads, with
$\tau_{\rm cent} = 0.1^{+0.7}_{-0.2}$ days,
which is consistent with zero lag and leaves
the question unanswered.

\begin{figure}
\begin{center}
    \leavevmode
  \centerline{\epsfig{file=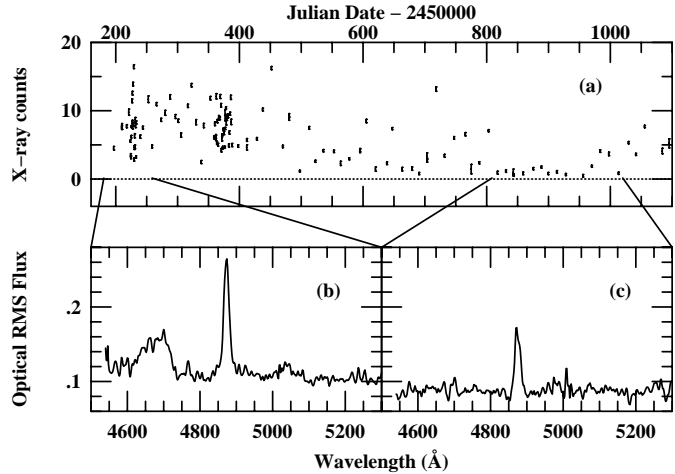,width=10cm,angle=0}}
  \end{center}
\caption{Comparison of hard X-ray and optical spectral
variations in NGC~4051 in different X-ray states.
Panel (a) shows the 2--10\,keV flux measured with
RXTE as a function time for the three-year
period 1996--1998.  Panel (b) shows the
rms optical spectrum during an X-ray active period in 1996,
Panel (c) shows the rms optical spectrum during an
X-ray quiescent period in 1998. Note the absence of strong
\heii\,$\lambda4686$ emission in panel (c). From
Peterson et al.$^{75}$ \copyright 2000 AAS.
\label{fig:adaf}}
\end{figure}

A final recent observation that should be mentioned
is based on a three-year combined X-ray (RXTE) and
optical monitoring program on the narrow-line
Seyfert 1 galaxy NGC 4051\cite{Pe_ea00}.
During the first two years
of this program, NGC 4051 behaved
in a ``normal'' fashion for narrow-line Seyfert 1s, 
with rapid, violent X-ray variability and much lower-amplitude
optical variations. During the third year, 
NGC~4051 went into a very low X-ray state.
In Fig.~\ref{fig:adaf}, the top panel shows
the X-ray light curve. The bottom two panels show
rms optical spectra, which isolate the variable parts of the 
spectrum,  obtained during the two periods indicated, one
during a high X-ray state and one during the low X-ray state.
They are remarkably different.
The high-state rms spectrum (lower left)
shows that the optical continuum and
the \Hbeta\ and \heii\,$\lambda4686$ emission lines were
all varying strongly. However, the low-state rms spectrum
(lower right) shows that the optical continuum and the
\Hbeta\ emission line are still present and variable, but the \heii\ line
has vanished from the rms spectrum, meaning that it is
absent or constant. Whether or not it has completely vanished
is difficult to determine on account of the blending of
\heii\ with strong \feii\ emission in the mean spectrum,
but an attempt to remove the \feii\ emission by subtraction
of a template indicates that most of the \heii\ emission
must have vanished during 1998. This indicates that
not only has the X-ray flux dramatically decreased, but
the EUV flux (which drives the \heii\ variations) must
have also decreased by a significant amount. This
may suggest that the entire inner accretion disk has 
undergone a transition to a low-radiation state, such
as an ``advection-dominated accretion flow'' \cite{Na_ea98}.

\subsection{Conclusion}

The few examples cited in this section give an indication
of the power of coordinated multiwavelength observations
for understanding the physical mechanisms at work in AGNs
on angular scales far too small to be resolvable with
any current or near-term technology. These are difficult
programs to implement, but the scientific return on them
is potential very large, especially as we can build on
the significant progress that has already been made.

\section*{Acknowledgments}
I am grateful for support of AGN variability studies
at The Ohio State University by the US National Science
Foundation through grant NSF-9420080 and by NASA
through LTSA grant NAG5-8397. I would like to thank
the other lecturers, especially S.\ Collin and H.\ Netzer, 
and my colleagues 
S.J.\ Collier,  S.\ Mathur,  R.W.\ Pogge,
J.C.\ Shields, and M.\ Vestergaard for suggestions on
the lectures and the manuscript. I am very grateful for
the hospitality extended to me at INAOE, particularly by
I.\ Aretxaga, J.\ Franco,  D.\ Kunth, and 
R.\ M\'{u}jica.

%
%
%
%

\end{document}